\documentclass[12pt,draftclsnofoot, onecolumn]{IEEEtran}

\usepackage{amsthm}
\usepackage{amsmath}
\usepackage{amssymb}
\usepackage{amsmath}
\usepackage{amssymb}
\usepackage{epsfig}
\usepackage{epsf}
\usepackage{amsmath,amsfonts,amsthm,bm} 
\usepackage{subfigure}
\usepackage{graphicx}
\usepackage{url}
\usepackage{color}
\usepackage[noend]{algorithmic}
\usepackage{algorithm}
\usepackage{verbatim}
\usepackage{environ}
\usepackage{mathtools}

\usepackage[numbers,sort&compress]{natbib}

\usepackage{enumitem,kantlipsum}

\usepackage{multicol}
\usepackage{tikz}
\usepackage{pgfplots}
\pgfplotsset{compat=1.7}
\usepgfplotslibrary{groupplots} 
\usetikzlibrary{pgfplots.groupplots}

\usepgfplotslibrary{fillbetween}

\usepackage{blindtext}

\usepackage{subfiles} 

\def\BibTeX{{\rm B\kern-.05em{\sc i\kern-.025em b}\kern-.08em
    T\kern-.1667em\lower.7ex\hbox{E}\kern-.125emX}}

\newtheorem{lemma}{Lemma}

\newtheorem{remark}{Remark}
\usepackage{url}

\include{defs}
\begin{document}
\title{



Control and Placement of Finite-Resolution Intelligent Surfaces in IoT Systems with Imperfect CSI
}


\author{Sajjad~Nassirpour,~Alireza~Vahid,~Dinh-Thuan~Do,~and~Dinesh~Bharadia

\thanks{S. Nassirpour, A. Vahid, and D. T. Do are with the Department of Electrical Engineering at the University of Colorado, Denver, USA. Emails: {\sffamily sajjad.nassirpour@ucdenver.edu}, {\sffamily alireza.vahid@ucdenver.edu}, and {\sffamily thuandinh.do@ucdenver.edu}.

D. Bharadia is with the Department of Electrical and Computer Engineering at the
University of California, San Diego, USA. Email: {\sffamily dineshb@eng.ucsd.edu}.}}

\maketitle


\begin{abstract}
In this paper, we study the advantages of using reconfigurable intelligent surfaces (RISs) for interference suppression in single-input single-output (SISO) distributed Internet of Things (IoT) networks. Implementing RIS-assisted networks confronts various problems, mostly related to the control and placement of the RIS. To tackle the control-related challenges, we consider noisy and local channel knowledge, based on which we devise algorithms to optimize the potentially distributed RISs to achieve an overall network objective, such as the sum-rate. We use a network with a centralized RIS as a benchmark for our comparisons. We further assume low-bit phase shifters at the RIS to capture real-world hardware limitations. We also study the placement of the RIS and analytically quantify the minimum required degrees-of-control for the RIS as a function of its location to guarantee a specific network performance metric and verify the results via simulations.
\end{abstract}


\section{Introduction}
\label{Section:Intro}
Internet of Things (IoT) communication networks enable devices with sensing, processing, and communication capabilities to provide a wide range of services, such as intelligent manufacturing, emergency detection, and structural health monitoring with minimal human interaction~\cite{mozaffari2017mobile}. Interference is an inevitable challenge in IoT networks due to the scarcity of the spectrum and the proximity of a large number of IoT users. This work studies a single-input single-output (SISO) IoT network with $K$ base stations (BSs) and $K$ IoT devices.


To overcome the aforementioned challenges,~\cite{ahsan2021resource,yang2019resource} use resource allocation techniques based on non-orthogonal multiple access (NOMA) to support IoT devices since IoT transmitters/receivers have limited size and battery capacity and are thus expected to have simple architectures. On the other hand, reconfigurable intelligent surface (RIS) has recently been proposed as a promising power-efficient alternative solution to handle interference~\cite{zhang2020capacity,yu2019miso,jiang2019over,guo2020weighted,fang2022optimum}. An RIS can be viewed as a set of programmable reflecting elements, each capable of altering the phase and possibly the amplitude of the incident signal, and is placed between the BSs and the receivers to modify and enhance the communication links. 


Intelligent surfaces propose an attractive solution in wireless networks as they are supposed to be (nearly-)passive and do not rely on traditional infrastructure~\cite{torkzaban2021shaping,kammoun2020asymptotic}. However, the implementation of RIS-assisted networks faces several challenges, mostly revolving around the control and placement of the intelligent surfaces. 


Controlling reconfigurable intelligent surfaces involves several steps, including channel estimation, dissemination of the channel knowledge throughout the network, optimizing the RIS configuration, and then altering the RIS configuration as desired. The presence of an RIS introduces a large number of new communication links, thus increasing the complexity of channel estimation~\cite{wang2020channel}. Next, the acquired channel knowledge needs to be disseminated throughout the network via feedback channels, which are typically rate-limited, delayed, or local~\cite{vahid2011interference,vahid2014capacity,vahid2019degrees,vahid2021erasure}. To enhance the existing literature, which mostly assumes noiseless and global channel knowledge~\cite{yu2020power,zhi2021two,zheng2021uplink,li2021double}, we study RIS-assisted networks with noisy and/or local knowledge of channel state information (CSI) and evaluate the resulting performance degradation. On the issue of RIS configuration and optimization, many prior results assume an ideal RIS that can alter phase continuously through its phase shifters (PSs). However, in reality, PSs have finite resolutions. In this work, we assume the latter model, which converts the continuous optimization into a discrete one and thus increases the complexity. We present an optimization method for this setting and compare it to relevant benchmarks to highlight its advantages. 


The placement of RIS has not been extensively studied. In most cases, RIS is either placed near the BSs or the users~\cite{yu2020power,li2021double,wu2021intelligent} to mitigate the impact of product path-loss~\cite{wu2021intelligent}. However, in practice, due to user mobility or physical constraints, neither may be feasible. To shed light in this direction, we quantify the minimum required degrees-of-control for the RIS (i.e., the number of RIS elements each with a known set of possible phase shifts) to attain a desired performance metric (e.g., per user rate) and analyze the results as a function of the location of the RIS. 


Our contributions are thus multi-fold. We study RIS-assisted networks with noisy and local, therefore imperfect, channel knowledge and quantify the impact of this limited knowledge on network performance. Further, we consider networks with {\it distributed} RISs and compare this setting to a benchmark with a centralized RIS. We present an optimization method for finite resolution (i.e., RIS with discrete PSs) and compare the results in terms of gain and complexity to relevant benchmarks. Additionally, we evaluate through simulations the RIS-assisted system performance under various fading models, including Nakagami, Rician, and Rayleigh. Finally, we provide design guidelines by quantifying the required number of RIS elements as a function of the location of the RIS(s) to attain certain overall network objectives, such as sum-rate or fairness, and verify the results through simulations.

The rest of the paper is structured as follows. We present our channel model based on centralized and distributed RISs and outline the optimization problem in Section~\ref{Section:Problem}. In Section~\ref{Section:opt_approach}, we discuss our proposed optimization approach. Then, in Section~\ref{Section:min_M}, we provide a lower bound on the minimum required degrees-of-control for the RIS. In Section~\ref{Section:Numerical}, we present the simulation results and then conclude the paper in Section~\ref{Section:conclusion}.
\subsection{Notations}
Throughout the paper, we use bold-face lowercase, bold-face uppercase, and italic letters to denote vectors, matrices, and scalars, respectively. We also use $\mathbb{C}^{I\times J}$ to describe the space of $I\times J$ complex-valued matrices. $\mathrm{diag}\{\textbf{x}\}$ is a diagonal matrix using vector $\textbf{x}$, and $|\textbf{x}|$ and $||\textbf{x}||$ are the absolute value and Euclidean norm of complex-valued vector $\textbf{x}$, respectively. 
We use $\mathbb{E}(.)$, $\log(.)$, $Re(.)$, and $Im(.)$ to denote the statistical expectation, logarithmic function in base 2, and real and imaginary parts of a complex number, respectively. Finally, $\textbf{B}^H$ shows the conjugate transpose of matrix $\textbf{B}$.
\section{Problem Setting}
\label{Section:Problem}
IoT systems should support a diverse set of users, some of which with very strict power and physical constraints. 
We thus focus on an IoT network 
where each transmitter and receiver is equipped with a single antenna
, and the communication is assisted by intelligent surfaces.
More specifically, we consider two RIS-based cases: (1) centralized RIS, which is useful when the RIS 
can be placed close to either the transmitters or the receivers, and (2) distributed RISs, where each transmitter has one dedicated RIS, which better mirrors the real-world challenges of IoT networks wherein each RIS could be installed on the building facade/wall close to its associated transmitter.

\begin{figure}[htb]
  \centering
  \includegraphics[trim = 0mm 0mm 0mm 0mm, clip, scale=9, width=0.95\linewidth, draft=false]{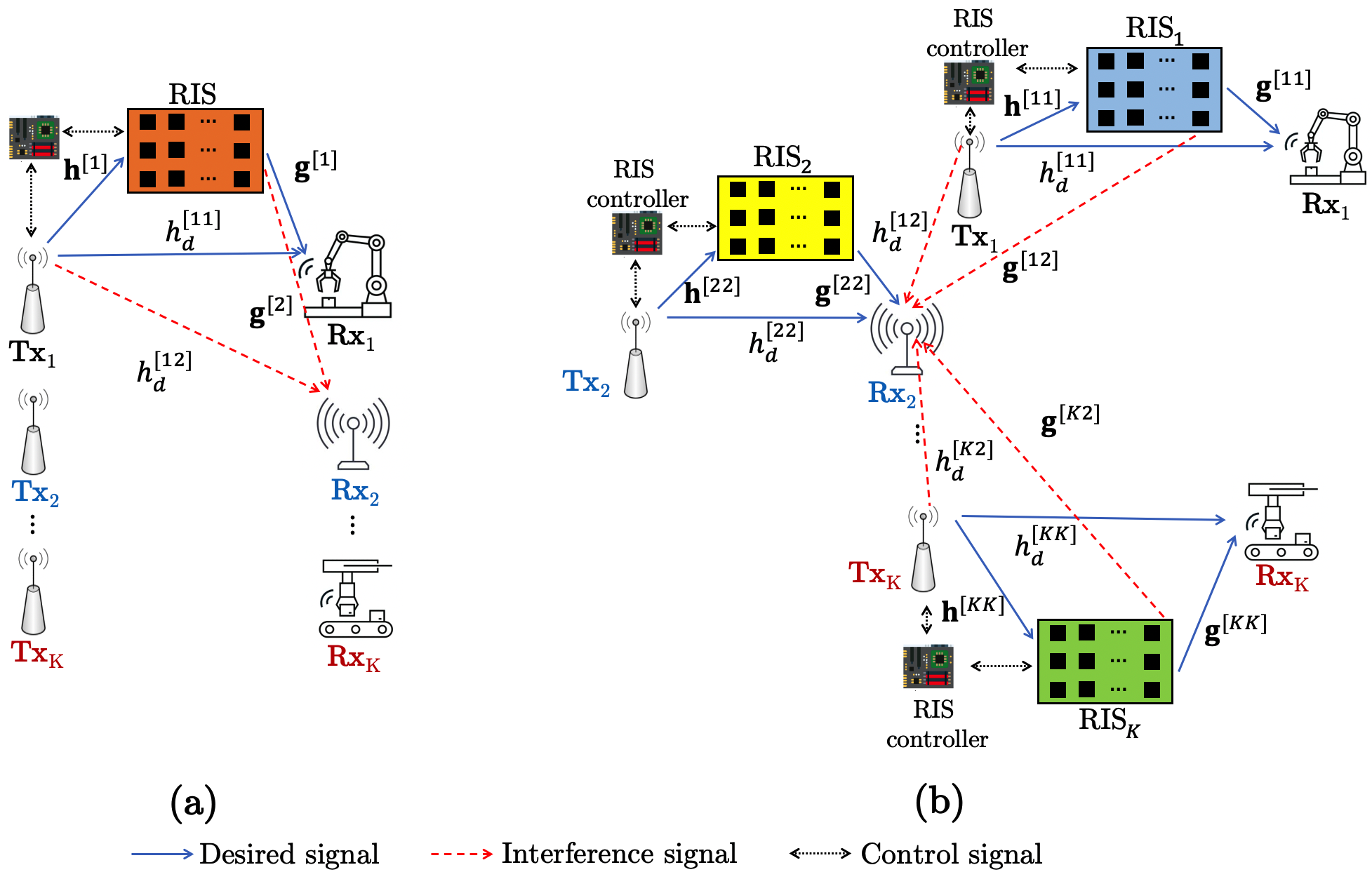}
  \vspace{-5mm}
  \caption{\it 
  A $K$-user 
  SISO IoT network with an $M$-element centralized RIS; 
  (b) A $K$-user SISO IoT network with $K$ distributed RISs where each RIS has $M$ elements.
  }\label{Fig:channel_model_RIS}
  \vspace{-7mm}
\end{figure}


\subsection{Centralized RIS}
\label{Subsection:channel_model_one_RIS}
In this case, we use a centralized RIS with $M$ reconfigurable elements
as shown in Fig.~\ref{Fig:channel_model_RIS}(a), where ${\sf Tx}_i$, RIS, and ${\sf Rx}_i$ are available at $(x^{[i]}_t,y^{[i]}_t)$, $(x_0,y_0)$, and $(x^{[i]}_r,y^{[i]}_r), i=1,2,\ldots,K$, respectively. 

\textbf{Channel model:}
We use $h_d^{[ji]}(t) \in \mathbb{C}^{1\times 1}$, $\textbf{h}^{[j]}(t) \in \mathbb{C}^{M\times 1}$, and $\textbf{g}^{[i]}(t) \in \mathbb{C}^{1\times M}$ to denote the channels of ${\sf Tx}_j$-${\sf Rx}_i$ link (between ${\sf Tx}_j$ and ${\sf Rx}_i, i, j \in \{1,2,\ldots , K\}$), ${\sf Tx}_j$-RIS link, and RIS-${\sf Rx}_i$ link, respectively, at time $t$. We assume RIS elements are spaced at least $\lambda/2$
apart from each other where $\lambda$ represents the wavelength of the transmitted signal. Therefore, the elements of $\textbf{h}^{[j]}(t)$ and $\textbf{g}^{[i]}(t)$
are independent across time and users. 
Later, we will describe the small-scale fading and large-scale fading to characterize the channels.

\textbf{Available CSI at transmitters:} 
Since the RIS elements need to be optimized to suppress interference at all receivers simultaneously, we assume global CSI is available at the transmitters~\cite{wu2019beamforming,wu2019intelligent,fu2021reconfigurable,bafghi2022degrees}. 
There is no further data exchange between the transmitters.

\textbf{Available CSI at receivers:} We assume each receiver is aware of global CSI since it requires global CSI to know the RIS configuration and subsequently decodes its message. We note that no data is exchanged beyond CSI between the receivers. 

Here, we assume each receiver estimates its incoming channels from all transmitters perfectly and then informs the others about these channels through two cases: (1) The noiseless links, which represent that other network nodes learn the channels as the receiver (noiseless CSI); (2) The noisy links, meaning that the information provided to the transmitters is noisy. We refer to this as ``noisy-$p$'' channels if other nodes attain the channels at a signal-to-noise ratio (SNR) of $p$ dB.   


To characterize the above communication channels, we consider the small-scale fading and large-scale fading as below:

\textbf{Small-scale fading:} Similar to~\cite{wu2019towards,papazafeiropoulos2021asymptotic,wu2018intelligent}, 
since in practice, RIS is usually
positioned with the knowledge of the BS’s location, we consider $\textbf{h}^{[j]}(t)$ is a line of sight (LoS) channel vector (i.e., $\textbf{h}^{[j]}(t)=\textbf{h}^{[j]}$). On the contrary, due to the user’s mobility and the
complex propagation environment, assuming $h_d^{[ji]}(t)$ and $\textbf{g}^{[i]}(t)$ are LoS channels is impossible in practice; thus, we assume $h_d^{[ji]}(t)$ and $\textbf{g}^{[i]}(t)$ are distributed based on Rayleigh fading. Here, we write $h_m^{[j]}$, the $m^{\mathrm{th}}$ element of $\textbf{h}^{[j]}$, as 
\begin{align}
    h_m^{[j]}=e^{-j\frac{2 \pi }{\lambda}d_{{\sf Tx}_j-\text{RIS}^{[m]}}},
\end{align}
where $d_{{\sf Tx}_j-\text{RIS}^{[m]}}$ shows the distance between ${\sf Tx}_j$ and the $m^{\mathrm{th}}$ element of the RIS. In addition, 
$h_d^{[ji]}(t)$ and $g_m^{[i]}(t)$ (the $m^{\mathrm{th}}$ element of $\textbf{g}^{[i]}(t)$) are distributed based on  $\mathcal{CN}(0,\sigma_{h_d}^2)$ and $\mathcal{CN}(0,\sigma_g^2)$, respectively, where $\sigma_{h_d}^2$ and $\sigma_g^2$ show the corresponding variances. 

We use a large array antenna to capture the RIS in this part. Later, in Section~\ref{Section:Numerical}, we will utilize metasurfaces to build the RIS and consider the angle between the incident and reflected signals at RIS. Further, we will evaluate the performance of our RIS-assisted network by considering Nakagami and Rician fading models as more realistic channel fading models in Section~\ref{Section:Numerical}. 

\textbf{Large-scale fading:} We use $C_0 d^{-\alpha}$ as the path loss profile where $C_0$ describes the signal loss at a reference distance (e.g., $1m$), $d$ is the distance between a pair of nodes (e.g., ${\sf Tx}_j$ and ${\sf Rx}_i$) and $\alpha$ represents the corresponding path loss exponent.

For simplicity, in the rest of the paper, we use $m_h$, $\sigma_{h_d}^2$, and $\sigma_g^2$ to denote $h_m^{[j]}$, the variance of $h_d^{[ji]}(t)$, and the variance of each element of $\textbf{g}^{[i]}(t)$, respectively, which include the impact of both small-scale fading and large-scale fading. 

\textbf{RIS configuration:} In this paper, we assume a controller between one of the transmitters (e.g., ${\sf Tx}_1$ in Fig.~\ref{Fig:channel_model_RIS}(a)) and the RIS to exchange CSI and control the RIS elements over a separate communication connection. Moreover, we consider only the first reflected signal from the RIS due to the significant path loss and ignore the hardware imperfection features (e.g., non-linearity) at the RIS; thus, if $\Tilde{x}(t)$ is the incident signal at time $t$, we find the reflected signal from the $m^{\mathrm{th}}$ element of the RIS as
\begin{align}
    \Tilde{y}_m\left(t\right)=\Tilde{x}\left(t\right)\left(\beta_m\left(t\right)e^{j\theta_m\left(t\right)}\right), \quad 1\leq m \leq M, 
\end{align}
where $\beta_m(t) \in[0,1]$ and $\theta_m(t) \in [0,2\pi)$ are the amplitude and phase of the $m^{\mathrm{th}}$ element of the RIS, respectively, at time $t$. Recently, \cite{nayeri2018reflectarray,wu2019towards,wu2019intelligent} have shown that it is hard to control $\beta_m(t)$ and $\theta_m(t)$ separately in the real-world scenarios; therefore, we consider $\beta_m(t)=1$ (i.e., perfect reflectors) at the RIS. 
We also assume that each RIS element has a $b$-bit PS, indicating that each RIS element can only take $N=2^b$ quantized levels. For simplicity, we assume these $N$ levels are distributed uniformly in $[0,2\pi)$. We define $\phi_N$, the set of quantized values of $\theta_m(t)$, as
\begin{align}
    \phi_N=\left\{0,\frac{2\pi}{N},\frac{4\pi}{N},\ldots,\frac{2\pi(N-1)}{N}\right\}.
\end{align}

Moreover, we use $\mathbf{\Theta}(t) \in \mathbb{C}^{M \times M}$ to denote the RIS configuration, which is given by
\begin{align}
    \mathbf{\Theta}(t)=\mathrm{diag}\{e^{\boldsymbol{\theta}(t)}\}, \quad \boldsymbol{\theta}(t)=[\theta_1(t),\theta_2(t),\ldots,\theta_{M}(t)].
\end{align}

\textbf{Received signal:} The received signal at ${\sf Rx}_i$ is a combination of the direct signals from the BSs and the reflected signals from the RIS, which is given by
\begin{align}
\label{eq:received_sig_a_RIS}
    y_i(t)=\sum_{j=1}^K \Big[\textbf{g}^{[i]}(t)\mathbf{\Theta}(t)\textbf{h}^{[j]}+h_d^{[ji]}(t)\Big]x_j(t)+n_i(t),
\end{align}
where $x_j(t)$ is the transmitted signal at ${\sf Tx}_j$, $n_i(t) \thicksim \mathcal{CN}(0,\sigma^2)$ 
is additive white Gaussian noise (AWGN) at ${\sf Rx}_i$, and $\mathbb{E}\{\textbf{x}(t)\textbf{x}^{H}(t)\}=\mathrm{diag}([P_1, P_2,\ldots,P_K])$ where $\textbf{x}(t)=[x_1(t),x_2(t),\ldots,$ $x_K(t)] \in \mathbb{C}^{1\times K}$ and $P_j, j\in \{1,2,\ldots K\}$ describes the transmit power from ${\sf Tx}_j$. As we describe later, all operations occur in a single coherence time; therefore, the time notation is removed throughout the rest of this work. Then, we use \eqref{eq:received_sig_a_RIS} and write the signal-to-interference-plus-noise ratio (SINR) at ${\sf Rx}_i$ as follows:
\begin{align}
\label{eq:SINR_update_a_RIS}
    \text{SINR}_i=\frac{P_i\Big|\textbf{g}^{[i]}\mathbf{\Theta}\textbf{h}^{[i]}+h_d^{[ii]}\Big|^2}{\sigma^2+\sum_{j=1, j\neq i}^{K }P_j\Big|\textbf{g}^{[i]}\mathbf{\Theta}\textbf{h}^{[j]}+h_d^{[ji]}\Big|^2}.
\end{align} 
\subsection{Distributed RISs}
As depicted in Fig.~\ref{Fig:channel_model_RIS}(b), we have $K$ distributed RISs, one dedicated RIS to each transmitter. We assume the same location notations for the transmitters and the receivers as discussed in Section~\ref{Subsection:channel_model_one_RIS} and use $(x^{[i]}_0,y^{[i]}_0)$ to denote the location of $\text{RIS}_i$, the $i^{\mathrm{th}}$ RIS.

\textbf{Channel model:} We use $\textbf{h}^{[ji]} \in \mathbb{C}^{M\times 1}$, and $\textbf{g}^{[ii]} \in \mathbb{C}^{1\times M}$ to denote the channels of ${\sf Tx}_j$-$\text{RIS}_i, i,j=1,2,\ldots,K$ and $\text{RIS}_i$-${\sf Rx}_i$ links, respectively. Notice that $h_d^{[ji]}$, large-scale fading, and small-scale fading are similar to the case with a centralized RIS. However, since $\text{RIS}_i$ is dedicated to ${\sf Tx}_i$, assuming LoS channels between ${\sf Tx}_j$ and $\text{RIS}_i$ is not feasible. As a result, we assume that $\textbf{h}^{[ji]}$ is distributed based on Rayleigh fading whose elements have $\mathcal{CN}(0,\Tilde{\sigma}_h^2)$ distribution.

\textbf{Available CSI at transmitters:} Acquiring global CSI requires an excessive overhead because of the substantial number of channels in the RIS-assisted network. This overhead would create a bottleneck in practice as feedback channels have limited bandwidth~\cite{vahid2010capacity,hu2012optimal,zhang2015adaptive,vahid2015impact,vahid2016two,vahid2017interference} and the delay overhead may render forward communications infeasible~\cite{vahid2021erasure,lin2021capacity,vahid2011interference}. 
Therefore, in distributed RIS case, we focus on a more realistic assumption as ${\sf Tx}_i$ is aware of its outgoing channels to the receivers (the direct channels and the channels through $\text{RIS}_i$) and beyond that, it only knows the statistics of the other channels. We refer to this model as local channel state information at the transmitters (local CSIT)
~\cite{vahid2010capacity,hu2012optimal,zhang2015adaptive,vahid2016two,vahid2017interference}. 

\textbf{Available CSI at receivers:} Here, ${\sf Rx}_i$ knows its incoming channels from all transmitters plus the outgoing channels from ${\sf Tx}_i$ to compute the optimal configuration at $\text{RIS}_i$ and beyond that, it only knows the statistics of the other channels.  

Similar to a centralized RIS, there is no data exchange between the transmitter and no data exchange beyond CSI between the receivers. Moreover, we have two channel knowledge cases as noiseless and noisy-$p$ channels.

\textbf{RIS configuration:} We assume each transmitter uses a controller to exchange CSI and control the PSs at its corresponding RIS via a separate communication link. Similar to the case with one RIS, we consider only the first reflected signal from each RIS in our calculations and exploit PS with $N$ possible levels at each RIS element, reflecting the signals perfectly. More precisely, we use $\theta^{[i]}_m \in \phi_N$ to denote the phase shift at the $m^{\mathrm{th}}$ element of $\text{RIS}_i$. Then, we use $\mathbf{\Theta}^{[i]} \in \mathbb{C}^{M \times M}$ to show the RIS configuration at $\text{RIS}_i$ as below.
\begin{align}
    \mathbf{\Theta}^{[i]}=\mathrm{diag}\{e^{\boldsymbol{\theta}^{[i]}}\}, \quad \boldsymbol{\theta}^{[i]}=[\theta^{[i]}_1,\theta^{[i]}_2,\ldots,\theta^{[i]}_{M}].
\end{align}

\textbf{Received signal:} According to the above, we find the received signal at ${\sf Rx}_i$, which is given by
\begin{align}
\label{eq:received_sig}
    y_i=\Big[\sum_{i^{\prime}=1}^K\textbf{g}^{[i^{\prime}i]}\mathbf{\Theta}^{[i^{\prime}]}\textbf{h}^{[ii^{\prime}]}+h_d^{[ii]}\Big]x_i+\sum_{j=1, j\neq i}^K \Big[\sum_{i^{\prime}=1}^K\textbf{g}^{[i^{\prime}i]}\mathbf{\Theta}^{[i^{\prime}]}\textbf{h}^{[ji^{\prime}]}+h_d^{[ji]}\Big]x_j+n_i.
\end{align}
Notice that the assumptions about $x_i$, transmit power $P_i$, and $n_i$ are similar to the assumptions described in \eqref{eq:received_sig_a_RIS}. Based on \eqref{eq:received_sig}, we compute the SINR at the $i^{\mathrm{th}}$ receiver as follows:
\begin{align}
\label{eq:SINR_update}
    \text{SINR}_i=\frac{P_i\Bigg|\sum_{i^{\prime}=1}^K\textbf{g}^{[i^{\prime}i]}\mathbf{\Theta}^{[i^{\prime}]}\textbf{h}^{[ii^{\prime}]}+h_d^{[ii]}\Bigg|^2}{\sigma^2+\sum_{j=1, j\neq i}^{K}P_j\Bigg|\sum_{i^{\prime}=1}^K\textbf{g}^{[i^{\prime}i]}\mathbf{\Theta}^{[i^{\prime}]}\textbf{h}^{[ji^{\prime}]}+h_d^{[ji]}\Bigg|^2}.
\end{align}

\subsection{Optimization Problem}
\label{Subsection:optimization_problem}

One of our goals in this work is to find the RIS configuration to maximize the sum-rate. Therefore, we define the following optimization problem with a centralized RIS.
\begin{align}
\label{eq:opt_min_one_RIS}
    &\underset{\boldsymbol{\theta}}{\max}\underbrace{\sum_{i=1}^{K} \log(1+\text{SINR}_i)}_{\overset{\triangle}=~-q(\boldsymbol{\theta})}\equiv \underset{\boldsymbol{\theta}}{\min}~q(\boldsymbol{\theta})
     \nonumber \\
    &\text{s.t.} ~~ \boldsymbol{\theta}=[\theta_1,\theta_2,\ldots,\theta_{M}], ~~ \theta_{m}\in \phi_N, \quad~ \text{for}~ 1\leq m \leq M.
\end{align}
Moreover, to meet fairness among users, we define a max-min optimization problem as
\begin{align}
\label{eq:opt_min_one_RIS_fairness}
    &\underset{\boldsymbol{\theta}}{\max}\underset{i=\{1,2,\ldots,K\}}{\min} \log(1+\text{SINR}_i) \nonumber \\
    &\text{s.t.} ~~ \boldsymbol{\theta}=[\theta_1,\theta_2,\ldots,\theta_{M}], ~~ \theta_{m}\in \phi_N, \quad~ \text{for}~ 1\leq m \leq M.
\end{align}
Notice that \eqref{eq:opt_min_one_RIS} and \eqref{eq:opt_min_one_RIS_fairness} are nonlinear integer programming (NLIP) problems, which are NP-hard. 
In general, there is no efficient way to find the exact solution to NLIP problems, and they can be solved through two methods: (i) converting it to an equivalent convex approximation problem and then applying a regular convex optimizer to find the approximated solution or (ii) using a heuristic method (e.g., filled function and genetic algorithm), which directly focuses on the original problem. In~\cite{wu2019beamforming,abdullah2022low}, the authors use the first case and study the successive refinement (SR) methods, which is an iterative algorithm that alternately optimizes each of the RIS elements by fixing the other $M-1$ elements in an iterative manner until convergence occurs. However, these methods get stuck in local solutions; therefore, in this paper, we utilize the second case and propose an optimization method based on a sigmoid filled function to find an approximation of the global solution. We will compare our approach with two SR-based methods in \cite{wu2019beamforming,abdullah2022low} as well as the simplified exhaustive search (SES) method and the genetic algorithm (GA), two well-known heuristic methods, in terms of complexity and rate in Section~\ref{Section:Numerical}.

Since only local CSIT is available to each transmitter in the distributed case, it is not feasible to maximize the sum-rate. Thus, in Section~\ref{subsect:main_result_dist_RIS}, we will define $score_i$ in \eqref{eq:score_i} for ${\sf Tx}_i$ and formulate the optimization problem accordingly.

%

\section{Proposed optimization approach}
\label{Section:opt_approach}

\subsection{Centralized RIS}
\label{subsect_optim_central_RIS}
The idea of the filled function optimization method was originally introduced in~\cite{renpu1990filled} for the continuous domain. Then,~\cite{ng2005discrete} updated it for the discrete domain. In this paper, we use a sigmoid-based filled function 
and run local and global searches to optimize \eqref{eq:opt_min_one_RIS}.

\textbf{Local search:}
We define $\mathcal{N}(\boldsymbol{\theta})$ as the neighbors of RIS configuration $\boldsymbol{\theta}$ such that
\begin{align}
    \mathcal{N}(\boldsymbol{\theta})=\boldsymbol{\theta} \cup \{\boldsymbol{\theta}+\mathbf{\bm{\lambda}}_m,\mathbf{\bm{\lambda}}_m\in \bm{\Lambda}\},
\end{align}
where $\mathbf{\bm{\lambda}}_m$ is an $M$-length column vector with the $m^\text{th}$ element chosen from $\{0,\frac{2\pi}{N},\frac{4\pi}{N},\ldots,$ $\frac{2\pi(N-1)}{N}\}$ and the others set to zero; further, $\bm{\Lambda}$ is the direction set equals to $\bm{\Lambda}=\{\mathbf{\bm{\lambda}}_m, m=1,2,\ldots, M\}$. Here, the local search scans all neighbors of $\boldsymbol{\theta}$ to find a solution as $\boldsymbol{\theta}^* \in \mathcal{N}(\boldsymbol{\theta})$ such that $q(\boldsymbol{\theta}^*) \leq q(\boldsymbol{\theta})$. If $\boldsymbol{\theta}^*=\boldsymbol{\theta}$, we call $\boldsymbol{\theta}^*$ the local minimizer; otherwise, we consider $\boldsymbol{\theta}=\boldsymbol{\theta}^*$ and repeat the local search. 


We provide Algorithm~\ref{algo:Local} to demonstrate how the local search works. 
\begin{algorithm}
  \caption{Local Search}
  \begin{multicols}{2}
  
  \hspace*{\algorithmicindent} \textbf{Input:} $\boldsymbol{\theta}_{\ell}$, $i_{\max}^{\text{loc}}$; \\ 
  \hspace*{\algorithmicindent} \textbf{Output:} $\boldsymbol{\theta}^*$;
  \begin{algorithmic}[1]
  \label{algo:Local}
  \STATE $i_{\ell}^{\text{loc}}=0$; $\boldsymbol{\theta}^*=\boldsymbol{\theta}_{\ell}$;
  \FOR{$\bm{\lambda}_m \in \bm{\Lambda}$}{
  \STATE $\Tilde{\boldsymbol{\theta}}=\boldsymbol{\theta}_{\ell}+\bm{\lambda}_m$;
  \IF{$q(\Tilde{\boldsymbol{\theta}})\leq q(\boldsymbol{\theta}^*)$}
  \STATE $\boldsymbol{\theta}^*=\Tilde{\boldsymbol{\theta}}$;
  \ENDIF
  }
  \ENDFOR
  \STATE $i_{\ell}^{\text{loc}} \leftarrow i_{\ell}^{\text{loc}}+1$;
  \IF{$\boldsymbol{\theta}^*\neq \boldsymbol{\theta}_{\ell}$ and $i_{\ell}^{\text{loc}}\leq i_{\max}^{\text{loc}}$}
  \STATE $\boldsymbol{\theta}_{\ell}=\boldsymbol{\theta}^*$;
  \STATE Go to line 2;
  \ELSE
  \STATE $\boldsymbol{\theta}^*$ is the local minimizer.
  \ENDIF
\end{algorithmic}
\end{multicols}
\end{algorithm}
We use $i_{\ell}^{\text{loc}}$ to denote the number of iterations to find the local optimizer in the $\ell^{\mathrm{th}}$ round of using Algorithm~\ref{algo:Local}. Then, we define $i_{\max}^{\text{loc}}$ as the maximum value of $i_{\ell}^{\text{loc}}$. Initially, we set $i_{\ell}^{\text{loc}}=0$ and $\boldsymbol{\theta}^*=\boldsymbol{\theta}_{\ell}$, and then, in lines 2 to 5, we check all the neighbors around $\boldsymbol{\theta}_{\ell}$ to find a better solution than $\boldsymbol{\theta}_{\ell}$. To do this, we set $\Tilde{\boldsymbol{\theta}}=\boldsymbol{\theta}_{\ell}+\bm{\lambda}_m$. If $q(\Tilde{\boldsymbol{\theta}})\leq q(\boldsymbol{\theta}^*)$, $\Tilde{\boldsymbol{\theta}}$ offers a better solution than $\boldsymbol{\theta}^*$; hence, we set $\boldsymbol{\theta}^*=\Tilde{\boldsymbol{\theta}}$. Following the completion of the search among $\boldsymbol{\theta}_{\ell}$'s neighbors, we increase $i_{\ell}^{\text{loc}}$ by one. If $\boldsymbol{\theta}^*=\boldsymbol{\theta}_{\ell}$ or $i_{\ell}^{\text{loc}}> i_{\max}^{\text{loc}}$, we consider $\boldsymbol{\theta}^*$ to be the local minimizer. Otherwise, we use $\boldsymbol{\theta}_{\ell}=\boldsymbol{\theta}^*$ and repeat lines $2$ to $5$ for all neighbors of new $\boldsymbol{\theta}_{\ell}$.

It is important to note that the local search uses finite steps to find the local minimizer since there are $N^M$ possible solutions that are finite. 

\textbf{Global search:} The filled function plays an essential role in the global search. In this work, we use sigmoid-based filled function $Q_r(\boldsymbol{\theta},\boldsymbol{\theta}^*)$ and define an auxiliary optimization problem as
\begin{align}
\label{eq:aux_opt}
    \underset{\boldsymbol{\theta}}{\min}~Q_r(\boldsymbol{\theta},\boldsymbol{\theta}^*) \nonumber \\
    \text{s.t.} \quad \boldsymbol{\theta},\boldsymbol{\theta}^*\in \phi_N,
\end{align}
where $\boldsymbol{\theta}^*$ is the current local minimizer of \eqref{eq:opt_min_one_RIS}, and $r>0$ denotes the filled function parameter. The aim here is to find a better solution than $\boldsymbol{\theta}^*$. 
The global search begins with random configuration $\boldsymbol{\theta}_0$ and uses the local search for \eqref{eq:opt_min_one_RIS} to obtain $\boldsymbol{\theta}^*_0$ as its local minimizer. Then, to find a better solution, it exploits a sigmoid filled function, which is given by 
\begin{align}
\label{eq:filled_func}
    Q_r(\boldsymbol{\theta},\boldsymbol{\theta}^*)=(1+\frac{1}{1+\beta||\boldsymbol{\theta}-\boldsymbol{\theta}^*||^2})f_r(q(\boldsymbol{\theta})-q(\boldsymbol{\theta}^*)),
\end{align}
where \begin{align}
\label{eq:parameters_filled_func}
f_r(q(\boldsymbol{\theta})-q(\boldsymbol{\theta}^*))=
\left\{ \begin{array}{ll}
\vspace{1mm} q(\boldsymbol{\theta})-q(\boldsymbol{\theta}^*)+r , & q(\boldsymbol{\theta})-q(\boldsymbol{\theta}^*)\leq -r, \\
\frac{1}{1+e^{\frac{-6}{r}(q(\boldsymbol{\theta})-q(\boldsymbol{\theta}^*)+r/2)}}, & -r<q(\boldsymbol{\theta})-q(\boldsymbol{\theta}^*)< 0,\\
1, & q(\boldsymbol{\theta})-q(\boldsymbol{\theta}^*)\geq 0,
\end{array} \right.
\end{align}
and
\begin{align}
    & \beta=
\left\{ \begin{array}{ll}
\vspace{1mm} 0, & q(\boldsymbol{\theta})-q(\boldsymbol{\theta}^*)\leq -r, \\
1, & \text{otherwise}.
\end{array} \right.
\end{align}
It applies $\boldsymbol{\theta}^*_0$ to \eqref{eq:filled_func} and runs the local search for \eqref{eq:aux_opt} to obtain $\Tilde{\boldsymbol{\theta}}^*_0$. Next, it assumes $\boldsymbol{\theta}_1=\Tilde{\boldsymbol{\theta}}^*_0$ and runs the local search for \eqref{eq:opt_min_one_RIS} with $\boldsymbol{\theta}_1$ to attain $\boldsymbol{\theta}^*_1$ as a new local minimizer. We call  $\boldsymbol{\theta}^*_1$ a better solution than the previous local minimizer if $q(\boldsymbol{\theta}^*_1)<q(\boldsymbol{\theta}^*_0)$. It repeats this procedure until a stopping criterion is satisfied. Here, we use $r$ as the optimization parameter. Intuitively, $r$ denotes a radius around $\boldsymbol{\theta}^*$ that the algorithm seeks for a new solution. If the global search fails to identify a better solution with $r$, it reduces $r$ by using $r=r/10$, resulting in a smaller search area surrounding the local minimizer. Notice that if the global search discovers a new solution, it resets $r$ to its initial value. Moreover, we use $i_{m}^{\text{filled}}, m=0,1,2,\ldots,(N-1)M $ to denote the number of times our method searches for a new solution by scanning the $m^{{\mathrm{th}}}$ neighbor of a local minimizer obtained from \eqref{eq:aux_opt}. Then, we define $i_{\max}^{\text{filled}}$ as the maximum number of times the global search uses the filled function to search for a new solution. Finally, we stop the procedure if $r<\epsilon$, for some  $\epsilon \ll 1$ (e.g., $\epsilon = 0.01$ in Table~\ref{table:rate_complexity_GS_methods})  or $\sum_{m=0}^{\left(N-1 \right)M} i_{m}^{\text{filled}}>i_{\max}^{\text{filled}}$, and declare the latest solution as the output of the global search. 

\begin{table} [htb]
\caption{A comparison between the original global search (OGS) method with the proposed global search (PGS) method in terms of sum-rate and complexity ratios when $K=4$, $\tau=10$, $r=10$, and $\epsilon=0.01$.}
\centering
\begin{tabular}{|c|c|c|c|c|}
\hline
Ratio & $M=4$ & $M=8$ & $M=16$ & $M=32$ \\[0.5ex]

\hline \hline
sum-rate (PGS)/sum-rate (OGS) & $1$ &  $1$ & $0.9985$ & $0.9974$ \\

complexity (PGS)/complexity (OGS) & $99\%$  &  $90\%$ & $84\%$ & $76\%$\\[1ex]  
\hline
\end{tabular}
\label{table:rate_complexity_GS_methods}
\end{table}
To accelerate the search process, in this paper, we modify the global search and use the local search to optimize \eqref{eq:opt_min_one_RIS} every $\tau \in \mathbb{N}$ times. More precisely, after $\tau$ times of finding the local minimizer of the auxiliary optimization problem, we run the local search for \eqref{eq:opt_min_one_RIS} to minimize $q(\boldsymbol{\theta})$. 
According to Table~\ref{table:rate_complexity_GS_methods}, the proposed global search achieves almost the same sum-rate as the original global search while reducing the complexity by $24 \%$ when $K=4$ and $M=32$.

We provide a pseudo code in Algorithm~\ref{algo:global} to describe the global search. The algorithm's inputs are $\boldsymbol{\theta}_0$, $r>1$, $\tau\geq 1$, $\epsilon\ll 1 $, $i_{\max}^{\text{loc}}$, and $i_{\max}^{\text{filled}}$, where $\boldsymbol{\theta}_0$ is selected randomly. Here, the output is $\boldsymbol{\theta}^{**}$, and we use $\ell,r_0$, and $\bar{\tau}$ to denote the $\ell^{\mathrm{th}}$ round of the global search, to keep the initial value of $r$, and to show which round of the global search uses the local search for \eqref{eq:opt_min_one_RIS}, respectively. Initially, we set $\ell=0$, $r_0=r$, $\bar{\tau}=\tau$, $i_{m}^{\text{filled}}=0, m=0,1,\ldots,(N-1)M$, and $\boldsymbol{\theta}^{**}=\boldsymbol{\theta}_0$. Then, if $\ell+1\geq \bar{\tau}$, we update $\bar{\tau}$ and run the local search for \eqref{eq:opt_min_one_RIS} with $\boldsymbol{\theta}_{\ell}$ to get its local minimizer as $\boldsymbol{\theta}^*_{\ell}$. Otherwise, we skip the local search and consider $\boldsymbol{\theta}^*_{\ell}=\boldsymbol{\theta}_{\ell}$. Next, if $q(\boldsymbol{\theta}^{*}_{\ell})<q(\boldsymbol{\theta}^{**})$ or $\ell=0$, we set $\boldsymbol{\theta}^{*}_{\ell}$ as a new optimal solution, reset the value of $r$ ($r=r_0$), and put $m=1$ where $m$ represents the index of a neighbor around $\boldsymbol{\theta}^*_{\ell}$. Then, we use the local search for \eqref{eq:aux_opt} with $\boldsymbol{\theta}^*_{\ell}$ to obtain $\bar{\boldsymbol{\theta}}^*_{\ell}$ and set $i_{m-1}^{\text{filled}} \leftarrow i_{m-1}^{\text{filled}}+1$. If $m=1$, we increase $\ell$ by one and set $\boldsymbol{\theta}_{\ell}=\bar{\boldsymbol{\theta}}^*_{\ell-1}$; else, the algorithm knows that it searched at least one of the neighbors around $\boldsymbol{\theta}^*_{\ell}$ and put $\boldsymbol{\theta}_{\ell}=\bar{\boldsymbol{\theta}}^*_{\ell}$. Then, it goes to line $4$. Thereafter, if $q(\boldsymbol{\theta}_{\ell})\geq q(\boldsymbol{\theta}^{**})$, the global search checks whether it covers all the neighbors around $\boldsymbol{\theta}^*_{\ell}$ or not. If the answer is no, it updates $\boldsymbol{\theta}^*_{\ell}$, increases $m$ by one, and goes to line $12$. Otherwise, if $r<\epsilon$ or $\sum_{m=0}^{(N-1)M}i_{m}^{\text{filled}}>i_{\max}^{\text{filled}}$, it stops the searching process and declares $\boldsymbol{\theta}^{**}$ as the global minimizer; else, it confines the searching area by reducing $r$. Then, it sets $\ell \leftarrow \ell-1$ and goes to line $11$. 
\begin{algorithm}
  \caption{Global Search}
  \begin{multicols}{2}
  \hspace*{\algorithmicindent} \textbf{Input:} $\boldsymbol{\theta}_0, r, \tau, \epsilon, i_{\max}^{\text{loc}}, i_{\max}^{\text{filled}}$; \\ 
  \hspace*{\algorithmicindent} \textbf{Output:} $\boldsymbol{\theta}^{**}$;
  \begin{algorithmic}[1]
  \label{algo:global}
  \STATE $\ell=0$; $r_0=r$; $\bar{\tau}=\tau$;
  \STATE $i_{m}^{\text{filled}}=0,$ for $m=0,1,\ldots, (N-1)M$;
  \STATE $\boldsymbol{\theta}^{**}=\boldsymbol{\theta}_0$;
  \IF{$\ell+1\geq \bar{\tau}$}
  \STATE $\bar{\tau}\leftarrow\bar{\tau}+\tau$;
  \STATE Run Algo.~\ref{algo:Local} with $\boldsymbol{\theta}_{\ell}$ and \eqref{eq:opt_min_one_RIS} to find $\boldsymbol{\theta}^*_{\ell}$;
  \ELSE
  \STATE  $\boldsymbol{\theta}^*_{\ell}=\boldsymbol{\theta}_{\ell}$;
  \ENDIF
  
  \IF{$q(\boldsymbol{\theta}^{*}_{\ell})<q(\boldsymbol{\theta}^{**})$ or $\ell=0$}
      \STATE $\boldsymbol{\theta}^{**}=\boldsymbol{\theta}^*_{\ell}$; $r=r_0$; 
      \STATE $m=1$;
      \STATE Run Algo.~\ref{algo:Local} with $\boldsymbol{\theta}^*_{\ell}$ and \eqref{eq:aux_opt} to get $\bar{\boldsymbol{\theta}}^*_{\ell}$;
      \STATE $i_{m-1}^{\text{filled}} \leftarrow i_{m-1}^{\text{filled}}+1$
      \IF{$m=1$}
      \STATE $\ell\leftarrow\ell+1$;
      \STATE $\boldsymbol{\theta}_{\ell}=\bar{\boldsymbol{\theta}}^*_{\ell-1}$;
      \ELSE
      \STATE $\boldsymbol{\theta}_{\ell}=\bar{\boldsymbol{\theta}}^*_{\ell}$;
      \ENDIF
      \STATE Go to line 4;
  \ENDIF
  \IF{$m \leq (N-1)M$}
      \STATE $\boldsymbol{\theta}^*_{\ell}=\boldsymbol{\theta}^*_{\ell-1}+ \bm{\lambda}_m$;
      \STATE $m\leftarrow m+1$;
      \STATE Go to line 12;
  \ELSE
      \IF{$r< \epsilon$ or $\sum_{m=0}^{(N-1)M}i_{m}^{\text{filled}}>i_{\max}^{\text{filled}}$}
           \STATE $\boldsymbol{\theta}^{**}$ is selected as global minimizer.
          \ELSE
          \STATE Reduce $r$ as $r \leftarrow \frac{r}{10}$;
          \STATE $\ell \leftarrow \ell-1$;
          \STATE Go to line 11;   
      \ENDIF
  \ENDIF
    \end{algorithmic}
    \end{multicols}
\end{algorithm}

\begin{remark}
We note that the filled function-based optimization methods can optimize a general nonlinear objective function as long as selecting a possible configuration from a discrete set is the only constraint. This model is usually referred to as an unconstrained optimization problem~\cite{ng2005discrete}.
\end{remark}
According to the above Remark, we can apply our sigmoid filled function method to the max-min optimization problem in \eqref{eq:opt_min_one_RIS_fairness}.
\subsection{Distributed RISs}
\label{subsect:main_result_dist_RIS}
In this case, we assume each transmitter only knows its own local CSIT; therefore, we define $score_i$ as a new objective function that captures the fact of enhancing the desired signals from ${\sf Tx}_i, i=1,2,\ldots,K$ at ${\sf Rx}_i$ while suppressing the interference signals from ${\sf Tx}_i$ at other receivers. Consequently, we write $score_i$ as
\begin{align}
\label{eq:score_i}
    score_i \triangleq\frac{P_i\Big|\textbf{g}^{[ii]}\mathbf{\Theta}^{[i]}\textbf{h}^{[ii]}+h_d^{[ii]}\Big|^2}{\sigma^2+\sum_{j=1, j\neq i}^{K}P_i\Big|\textbf{g}^{[ij]}\mathbf{\Theta}^{[i]}\textbf{h}^{[ii]}+h_d^{[ij]}\Big|^2},
\end{align}
where compared to $\text{SINR}_i$ given in \eqref{eq:SINR_update_a_RIS}, the denominator includes noise and the interference caused by ${\sf Tx}_i$ at unintended receivers.
\begin{figure}[ht]
  \centering
  \includegraphics[trim = 0mm 0mm 0mm 0mm, clip, scale=7, width=0.99\linewidth, draft=false]{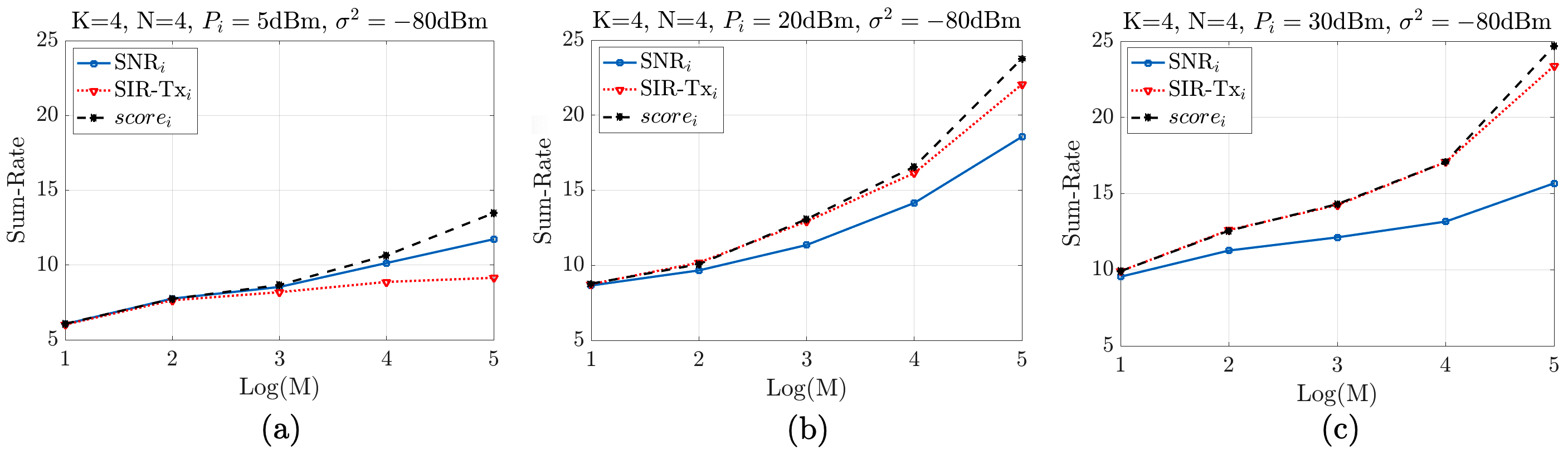}
  \vspace{-5mm}
  \caption{\it 
  Sum-rate versus $\log(M)$ with different objective functions as $\text{SNR}_i$, SIR-${\sf Tx}_i$, and $score_i$.
  We assume 
  $K=4$, $N=4$, $\sigma^2=-80$dBm, and $P_i \in \{5, 20, 30\}$ in dBm scale.} 
  \label{Fig:sinr_sir_snr}
\end{figure}
Then, we define the optimization problem as follows:
\begin{align}
\label{eq:opt_max_min_score}
    &\underset{\boldsymbol{\theta}^{[i]}}{\max}~score_i \nonumber \\
    &\text{s.t.} ~~ \boldsymbol{\theta}^{[i]}=[\theta^{[i]}_1,\theta^{[i]}_2,\ldots,\theta^{[i]}_{M}], ~~ \theta^{[i]}_{m}\in \phi_N, \quad~ \text{for}~ 1\leq m \leq M, ~ 1\leq i \leq K.
\end{align}

Here, we use the same strategy as Section~\ref{subsect_optim_central_RIS} to maximize \eqref{eq:opt_max_min_score} since the filled function can be applied to a general unconstrained optimization problem.

Notice that it is possible to use two other methods to define $score_i$: (1) SNR at the $i^{\mathrm{th}}$ receiver ($\text{SNR}_i$), and (2) signal-to-interference ratio from ${\sf Tx}_i$ (SIR-${\sf Tx}_i$) by removing noise from the denominator of \eqref{eq:score_i}. In Fig.~\ref{Fig:sinr_sir_snr}, we compare the performance of these two alternatives with $score_i$ in \eqref{eq:score_i} utilizing different transmit powers. It depicts that $score_i$ consistently outperforms the other two ways 
since it includes the impact of desired and interference signals from ${\sf Tx}_i$ as well as noise. In addition, Fig.~\ref{Fig:sinr_sir_snr} shows that the sum-rate corresponding to $\text{SNR}_i$ is higher than the sum-rate corresponding to SIR-${\sf Tx}_i$ when $P_i$ is small because the desired and interference signals are weak. However, SIR-${\sf Tx}_i$ outperforms $\text{SNR}_i$ at high $P_i$ values due to the strong interference signals. Fig.~\ref{Fig:sinr_sir_snr} also depicts that the difference between the curves using $score_i$ and SIR-${\sf Tx}_i$ decreases as $P_i$ increases.
\section{Minimum required degrees-of-control for the RIS}
\label{Section:min_M}
Prior results, for the most part, either do not consider the number of RIS elements and instead focus on adjusting the elements of a given RIS~\cite{wu2019beamforming,dunna2020scattermimo} or obtain the number of RIS elements at high SNRs~\cite{bafghi2022degrees}. In particular,~\cite{dunna2020scattermimo} uses a given RIS and configures its elements to match the received power from the reflected path to that of the direct path. In~\cite{wu2019beamforming}, the authors optimize the available RIS elements to minimize the transmitted power. Finally,~\cite{bafghi2022degrees} shows that $K(K-1)$ RIS elements are required to achieve DoF of $K$ using an active RIS and assuming $\text{SNR} \to \infty$. On the contrary, in this work, we compute a lower bound on the minimum number of RIS elements to guarantee to achieve a specific value of $\text{SINR}$ at each receiver. 
\subsection{Centralized RIS}
\label{subsect:M_min_one_RIS}

Usually, RIS elements are configured based on channel coefficients; however, to find a lower bound on $M$ in this work, we assume that $\boldsymbol{\theta}$ is independent of the channels and selected randomly with uniform distribution. For simplicity, we focus on a symmetric scenario, which includes the following: (i) All transmitters have the same transmit power (i.e., $P_i=P, i \in \{1,2,,\ldots K\}$); Each receiver stands at the same distance from all transmitters; (iii) The distance between the RIS and all transmitters are the same; (iv) The receivers are placed at the same distance from the RIS. We use the following two lemmas to explain our analysis. 


\begin{lemma}
\label{lemma:avg_variance}
If $\textbf{g}^{[i]}$ shows the channel vector between RIS and ${\sf Rx}_i$, whose elements are independent and identically distributed (i.i.d.) random variables with zero mean and variance $\sigma_{g}^2$, and  $\textbf{h}^{[j]}$ denotes the LoS channel vector between ${\sf Tx}_j$ and RIS with value $m_h$. Then, for a given RIS configuration $\mathbf{\Theta}$, where all elements are drawn independently at random from $\{0,\frac{2\pi}{N},\frac{4\pi}{N},\ldots,\frac{2\pi(N-1)}{N}\}$, the variance of $\textbf{g}^{[i]}\mathbf{\Theta}\textbf{h}^{[j]}$ is equal to
\begin{align}
    \label{eq:lemma_avg_variance}
    \nu=M\sigma_{g}^2(m_h)^2.
\end{align}
\end{lemma}
\begin{proof}
We have 
\begin{align}
\label{eq:proof_variance_lemma}
    \nu&
    =\text{var}\left(\sum_{m=1}^Mg_{m}^{[i]}e^{j\theta_m}h_{m}^{[j]}\right)\overset{(a)}{=}\sum_{m=1}^M\text{var}\left(g_{m}^{[i]}e^{j\theta_m}\right)\left(m_h\right)^2 \nonumber \\
    &\overset{(b)}{=}\sum_{m=1}^M\left(\sigma_{g}^2\underbrace{\text{var}\left(e^{j\theta_m}\right)}_{1}+\sigma_{g}^2\underbrace{\left[\mathbb{E}\left(e^{j\theta_m}\right)\right]^2}_{=0}+\underbrace{\left[\mathbb{E}\left(g_{m}^{[i]}\right)\right]^2}_{=0}\text{var}\left(e^{j\theta_m}\right)\right)\left(m_h\right)^2=M\sigma_{g}^2\left(m_h\right)^2,
\end{align}
where $\text{var}(X)$ denotes the variance of $X$. Here, $(a)$ holds since $h_{m}^{[j]}$ is constant and $e^{j\theta_m}$ and $g_{m}^{[i]}$ are independent. Further, $(b)$ follows the rule of the variance of the product of two independent variables. As a result, \eqref{eq:proof_variance_lemma} completes the proof.
\end{proof}

Then, we have
\begin{lemma}
\label{lemma:approx_prob}
For any $\delta_{ji}>0$, 
\begin{align}
\label{eq:prob_less_delta}
    \Pr\Big(\Big|\textbf{g}^{[i]}\mathbf{\Theta}\textbf{h}^{[j]}\Big|<\frac{\delta_{ji}}{2}\Big|\mathbf{\Theta}\Big)  \nonumber&>\Pr\Big(\Big|Re\{\textbf{g}^{[i]}\mathbf{\Theta}\textbf{h}^{[j]}\}\Big|<\frac{\delta_{ji}}{2\sqrt{2}}\Big|\mathbf{\Theta}\Big)\Pr\Big(\Big|Im\{\textbf{g}^{[i]}\mathbf{\Theta}\textbf{h}^{[j]}\}\Big|<\frac{\delta_{ji}}{2\sqrt{2}}\Big|\mathbf{\Theta}\Big)  \nonumber \\
    &\approx(\frac{\delta_{ji}}{\sqrt{2\pi \nu}})^2, \quad \quad i,j\in\{1,2,\ldots,K\}, j\neq i.
\end{align}
\end{lemma}
\begin{proof}
We defer the proof to Appendix~\ref{appndx:proof_approx_prob}.
\end{proof}
Similar to Lemma~\ref{lemma:approx_prob}, for any $\delta_{ji}>0$, 
\begin{align}
\label{eq:prob_less_delta_hd}
    \Pr\Big(\Big|h_d^{[ji]}\Big|<\frac{\delta_{ji}}{2}\Big|\mathbf{\Theta}\Big)&>\Pr\Big(\Big|Re\{h_d^{[ji]}\}\Big|<\frac{\delta_{ji}}{2\sqrt{2}}\Big|\mathbf{\Theta}\Big)\Pr\Big(\Big|Im\{h_d^{[ji]}\}\Big|<\frac{\delta_{ji}}{2\sqrt{2}}\Big|\mathbf{\Theta}\Big) \nonumber \\
    & \approx(\frac{\delta_{ji}}{\sigma_{h_d}\sqrt{2\pi }})^2, \quad \quad i,j\in\{1,2,\ldots,K\}, j\neq i.
\end{align}

Further, for any $\Delta_{ii}>0$, we have 
\begin{align}
\label{eq:greater_Delta}
    \Pr\Big(\Big|h_d^{[ii]}\Big|>\Delta_{ii}\Big|\mathbf{\Theta}\Big)&>\Pr\Big(\Big|Re\{h_d^{[ii]}\}\Big|>\frac{\Delta_{ii}}{\sqrt{2}}\Big|\mathbf{\Theta}\Big)\Pr\Big(\Big|Im\{h_d^{[ii]}\}\Big|>\frac{\Delta_{ii}}{\sqrt{2}}\Big|\mathbf{\Theta}\Big)  \nonumber \\
    &=\Big[2\int_{\frac{\Delta_{ii}}{\sqrt{2}}}^{\infty}\frac{1}{\sigma_{h_d}\sqrt{\pi }}e^{\frac{-r^2}{\sigma^2_{h_d}}}dr\Big]^2=[2Q(\frac{\Delta_{ii}}{\sigma_{h_d}})]^2, \quad \quad i\in\{1,2,\ldots,K\},
\end{align}
where $Q(.)$ represents Q-function. 
Finally, we present a lower bound on the minimum number of RIS elements as follows.
\begin{lemma}
\label{lemma:Min_M_RIS}
To achieve a per-user given value of SINR as $\tilde{\text{SINR}}$, the minimum number of RIS elements should follow:
\begin{align}
    M\geq\underset{a}{\min}\Big\{K^2\log_N\Big[\frac{\Tilde{\text{SINR}} 2\pi (K-1)(512\nu^{\prime})}{\left(\sigma_{h_d}\Big[Q^{-1}\big(\frac{1}{2N^{a/2K}}\big)\Big]-N^{\frac{-(10-a)}{2K^2}}\sqrt{\frac{\pi}{2} 512\nu^{\prime}}\right)^2-\frac{\Tilde{\text{SINR}}\sigma^2}{P}}\Big]+a\Big\},
\end{align}
where $Q^{-1}(.)$ is the inverse Q-function, $\log_N$ is a logarithmic function in base $N$, $\nu^{\prime}=\nu/M$, and $a$ shows a trade-off between signal enhancement from the desired transmitter and suppressing the interference signals from other transmitters. For instance, we use $a=0$ to dedicate RIS elements to reduce the interference and $a=M$ to exploit RIS elements to improve the desired signal. 
\end{lemma}
\begin{proof}
Based on \eqref{eq:SINR_update_a_RIS}, to calculate $\text{SINR}_i$, we need to know the values of $\Big|\textbf{g}^{[i]}\mathbf{\Theta}\textbf{h}^{[j]}+h_d^{[ji]}\Big|$ and $\Big|\textbf{g}^{[i]}\mathbf{\Theta}\textbf{h}^{[i]}+h_d^{[ii]}\Big|$. Here, we aim to have $\Big|\textbf{g}^{[i]}\mathbf{\Theta}\textbf{h}^{[j]}+h_d^{[ji]}\Big|<\delta_{ji}$ and $\Big|\textbf{g}^{[i]}\mathbf{\Theta}\textbf{h}^{[i]}+h_d^{[ii]}\Big|>\Delta_{ii}$ at ${\sf Rx}_i$. For simplicity, we use $\delta_{ji}=\delta$ and $\Delta_{ii}=\Delta$. Then, we have
\begin{align}
\label{eq:inequal_delta}
    \Big|\textbf{g}^{[i]}\mathbf{\Theta}\textbf{h}^{[j]}+h_d^{[ji]}\Big|\overset{(a)}{\leq} \Big|\textbf{g}^{[i]}\mathbf{\Theta}\textbf{h}^{[j]}\Big|+\Big|h_d^{[ji]}\Big|<\delta,
\end{align}
where $(a)$ follows the triangle inequality. Further, $\Big|\textbf{g}^{[i]}\mathbf{\Theta}\textbf{h}^{[i]}+h_d^{[ii]}\Big|$ can be lower-bounded by
\begin{align}
\label{eq:inequal_Delta}
    \Big|\textbf{g}^{[i]}\mathbf{\Theta}\textbf{h}^{[i]}+h_d^{[ii]}\Big|\geq \Big|h_d^{[ii]}\Big|-\Big|\textbf{g}^{[i]}\mathbf{\Theta}\textbf{h}^{[i]}\Big|>\Delta.
\end{align}
The inequalities in \eqref{eq:inequal_delta} and \eqref{eq:inequal_Delta} can be met if $\Big|\textbf{g}^{[i]}\mathbf{\Theta}\textbf{h}^{[j]}\Big|<\delta/2$, $\Big|h_d^{[ji]}\Big|<\delta/2$, and $\Big|h_d^{[ii]}\Big|>\Delta+\delta/2$. We calculate the probability of these three events using  \eqref{eq:prob_less_delta}, \eqref{eq:prob_less_delta_hd}, and \eqref{eq:greater_Delta}, respectively. In addition, there are $N^M$ different RIS configurations. Therefore, to have one RIS configuration that satisfies \eqref{eq:inequal_delta} and \eqref{eq:inequal_Delta} from all transmitters at all receivers, we need
\begin{align}
    \label{eq:proof_min_M_ini}
    N^{M}[2Q(\frac{\Delta+\delta/2}{\sigma_{h_d}})]^{2K}(\frac{\delta}{\sqrt{2 \pi M \nu^{\prime}}})^{2K}(\frac{\delta}{\sigma_{h_d}\sqrt{2 \pi }})^{2K(K-1)}>1.
\end{align}
Without loss of generality, we assume $M \nu^{\prime}\geq\sigma^2_{h_d}$, which happens if the reflected signal through the RIS is stronger than the signal from the direct path. Then, we rewrite \eqref{eq:proof_min_M_ini} as 
\begin{align}
    \label{eq:proof_min_M}
    N^{M}[2Q(\frac{\Delta+\delta/2}{\sigma_{h_d}})]^{2K}(\frac{\delta}{\sqrt{2 \pi M \nu^{\prime}}})^{2K^2}>1.
\end{align}

To calculate \eqref{eq:proof_min_M}, we assume
\begin{align}
\label{eq:assumption_delta}
    \delta=N^{\frac{-(M-a)}{2K^2}}\sqrt{2 \pi M^{+}\nu^{\prime}},
\end{align}
where $M^{+}$ is the maximum value of $M$ and $0 \leq a \leq M$ illustrates a trade-off between improving the desired signal and suppressing the interference signals. Then, by substituting $\delta$ in \eqref{eq:proof_min_M}, we obtain
\begin{align}
    N^a[2Q(\frac{\Delta+\delta/2}{\sigma_{h_d}})]^{2K}(\frac{M^{+}}{M})^{K^2}>1,
\end{align}
where $(\frac{M^{+}}{M})^{K^2}>1$ because $M<{M^{+}}$. Thus, we need
\begin{align}
    Q(\frac{\Delta+\delta/2}{\sigma_{h_d}})=\frac{1}{2N^{a/2K}},
\end{align}
to meet the inequality in \eqref{eq:proof_min_M}. As a result, we calculate $\Delta$ as
\begin{align}
\label{eq:Delta_val}
    \Delta&=\sigma_{h_d}\Big[Q^{-1}\big(\frac{1}{2N^{a/2K}}\big)\Big]-\frac{\delta}{2} \nonumber \\
    &\overset{\eqref{eq:assumption_delta}}{=}\sigma_{h_d}\Big[Q^{-1}\big(\frac{1}{2N^{a/2K}}\big)\Big]-N^{\frac{-(M-a)}{2K^2}}\sqrt{\frac{\pi}{2} M^{+}\nu^{\prime}}.
\end{align}

Now, we use \eqref{eq:SINR_update_a_RIS} to compute $\text{SINR}$ at ${\sf Rx}_i$ as below: 
\begin{align}
\label{eq:sinr_tild_0}
    \text{SINR}_i&=\frac{P\Big|\textbf{g}^{[i]}\mathbf{\Theta}\textbf{h}^{[i]}+h_d^{[ii]}\Big|^2}{\sigma^2+P\sum_{j=1, j\neq i}^{K }\Big|\textbf{g}^{[i]}\mathbf{\Theta}\textbf{h}^{[j]}+h_d^{[ji]}\Big|^2}\overset{(b)}{>}\frac{P\Delta^2}{\sigma^2+P(K-1)\delta^2}  \nonumber \\
    &=\frac{\left(\sigma_{h_d}\Big[Q^{-1}\big(\frac{1}{2N^{a/2K}}\big)\Big]-N^{\frac{-(M-a)}{2K^2}}\sqrt{\frac{\pi}{2} M^{+}\nu^{\prime}}\right)^2}{\frac{\sigma^2}{P}+2 \pi (K-1)N^{\frac{-(M-a)}{K^2}}(M^{+}\nu^{\prime})} \nonumber \\
    &\overset{(c)}{>}\frac{\left(\sigma_{h_d}\Big[Q^{-1}\big(\frac{1}{2N^{a/2K}}\big)\Big]-N^{\frac{-(M^{-}-a)}{2K^2}}\sqrt{\frac{\pi}{2} M^{+}\nu^{\prime}}\right)^2}{\frac{\sigma^2}{P}+2 \pi (K-1)N^{\frac{-(M-a)}{K^2}}(M^{+}\nu^{\prime})},
\end{align}
where $(b)$ holds using \eqref{eq:inequal_delta} and \eqref{eq:inequal_Delta}, and $(c)$ is correct by assuming
$M^{-}$ as the minimum value of $M$. Our goal is to attain a lower bound on $M$ that ensures $\text{SINR}_i>\Tilde{\text{SINR}}$. To do so, the right-hand side of \eqref{eq:sinr_tild_0} should be greater than or equal to $\Tilde{\text{SINR}}$. Consequently, we have
\begin{align}
\label{eq:sinr_tild}
   \frac{\left(\sigma_{h_d}\Big[Q^{-1}\big(\frac{1}{2N^{a/2K}}\big)\Big]-N^{\frac{-(M^{-}-a)}{2K^2}}\sqrt{\frac{\pi}{2} M^{+}\nu^{\prime}}\right)^2}{\frac{\sigma^2}{P}+2 \pi (K-1)N^{\frac{-(M-a)}{K^2}}(M^{+}\nu^{\prime})} \geq \Tilde{\text{SINR}}.
\end{align}
We note that \eqref{eq:sinr_tild} depends on the value of $a$; therefore, by taking $\log_N$ from both sides of \eqref{eq:sinr_tild} and considering $10 <M< 512$, we have
\begin{align}
\label{eq:final_lower_bound_equation}
    M\geq\underset{a}{\min}\Big\{K^2\log_N\Big[\frac{\Tilde{\text{SINR}} 2\pi (K-1)(512\nu^{\prime})}{\left(\sigma_{h_d}\Big[Q^{-1}\big(\frac{1}{2N^{a/2K}}\big)\Big]-N^{\frac{-(10-a)}{2K^2}}\sqrt{\frac{\pi}{2} 512\nu^{\prime}}\right)^2-\frac{\Tilde{\text{SINR}}\sigma^2}{P}}\Big]+a\Big\},
\end{align}
which completes the proof.
\end{proof}
Later, in Section~\ref{Section:Numerical}, we calculate the optimal value of $a$ that maximizes the left-hand side of \eqref{eq:sinr_tild} numerically. We define $M_{\min}$ as the smallest integer that satisfies \eqref{eq:final_lower_bound_equation}. Further, we refer to the curve that shows the average sum-rate versus different values of $P$ using $M_{\min}$ RIS elements as the theoretical-bound. 

\subsection{Distributed RISs}
In this part, we use the following lemma to show a lower bound on $M$ that guarantees achieving a given value of $score_i$ as $\Tilde{score}$ in a symmetric scenario as described in  Section~\ref{subsect:M_min_one_RIS}.

\begin{lemma}
\label{lemma:Min_M_score}
To achieve a per-user given value of $\text{score}_i$, as $\tilde{\text{score}}$, the minimum number of RIS elements at $\text{RIS}_i$, where $10<M<512$, should follow:
 \begin{align}
    M>\underset{a}{\min}\Big\{(K-1)\log_N\Big[\frac{\Tilde{score}\frac{\pi}{2}(K-1)(\sigma_{h_d}^2+512\nu^{\prime})}{(\sigma_{h_d}^2+10\nu^{\prime})\Big[Q^{-1}\big(\frac{1}{2N^{a/2}}\big)\Big]^2-\frac{\sigma^2}{P}\Tilde{score}}\Big]+a\Big\},
\end{align}
\end{lemma}

\begin{proof}
The proof is similar to Lemma~\ref{lemma:Min_M_RIS}.
\end{proof}
\section{Simulation results}
\label{Section:Numerical}
This section presents the numerical results of our proposed method, which are performed in MATLAB. 
The simulation results are averaged across $100$ Monte-Carlo trials. We present the results in two parts. First, we compare distributed RISs with a centralized RIS when the channels follow the small-scale fading outlined in Section~\ref{Section:Problem}. More precisely, we study the sum-rate and show the outage capacity using these two scenarios. In the second part, we assess the performance of our filled function optimization method with different benchmarks using a centralized RIS. In particular, we consider four baselines: the SES method, GA, SR~\cite{wu2019beamforming} method, and Modified SR (M-SR)~\cite{abdullah2022low} method. We compare our method with them in terms of rate (sum-rate and minimum rate) and complexity when the channels between the BSs and RIS are distributed based on the Rician fading model. Then, we utilize a more realistic channel model where all channels follow Nakagami fading model. Finally, we study the minimum required degrees-of-control for the RIS. Specifically, we consider a centralized RIS, compare the theoretical-bound with the simulation findings, and obtain the minimum number of RIS elements to provide a certain sum-rate when the RIS moves from the BSs toward the users.  


Throughout this section, we set $\sigma^2=-80$dBm, $\tau=10,r=10$, and $\epsilon=0.01$, and we consider $3.5$, $2$, and $2.1$ as the path loss exponents between the transmitters and the receivers, between the transmitters and the RIS(s), and between the RIS(s) and the receivers, respectively. 
\subsection{Distributed RISs vs. centralized RIS}

\begin{figure}[ht]
\vspace{-8mm}
	\begin{minipage}[c][1\width]{
	   0.32\textwidth}
	   \centering
	   \includegraphics[width=1.1\textwidth]{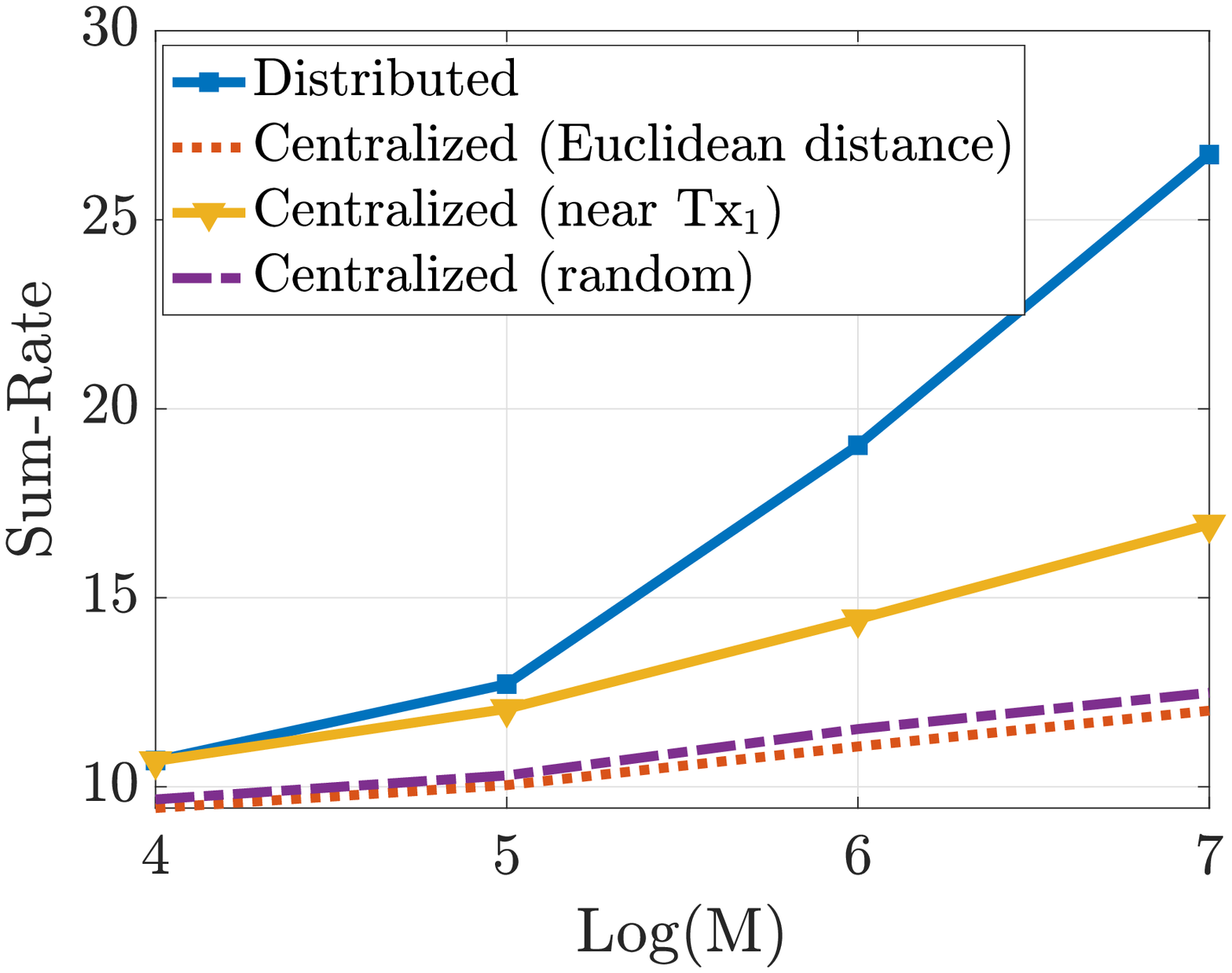}
	\end{minipage}
	\begin{minipage}[c][1\width]{
	   0.32\textwidth}
	   \centering
	   \includegraphics[width=1.1\textwidth]{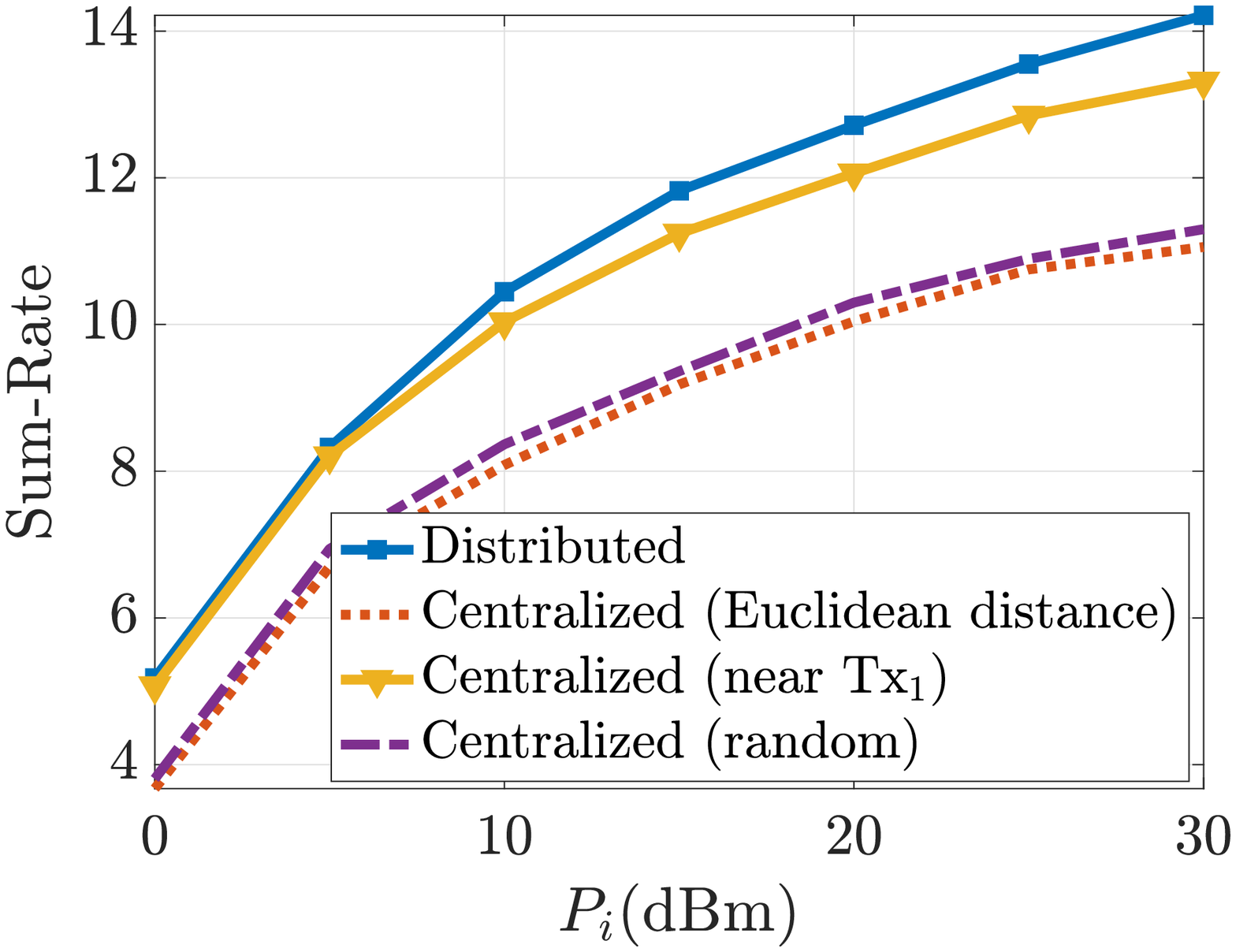}
	\end{minipage}
	\begin{minipage}[c][1\width]{
	   0.32\textwidth}
	   \centering
	   \includegraphics[width=1.1\textwidth]{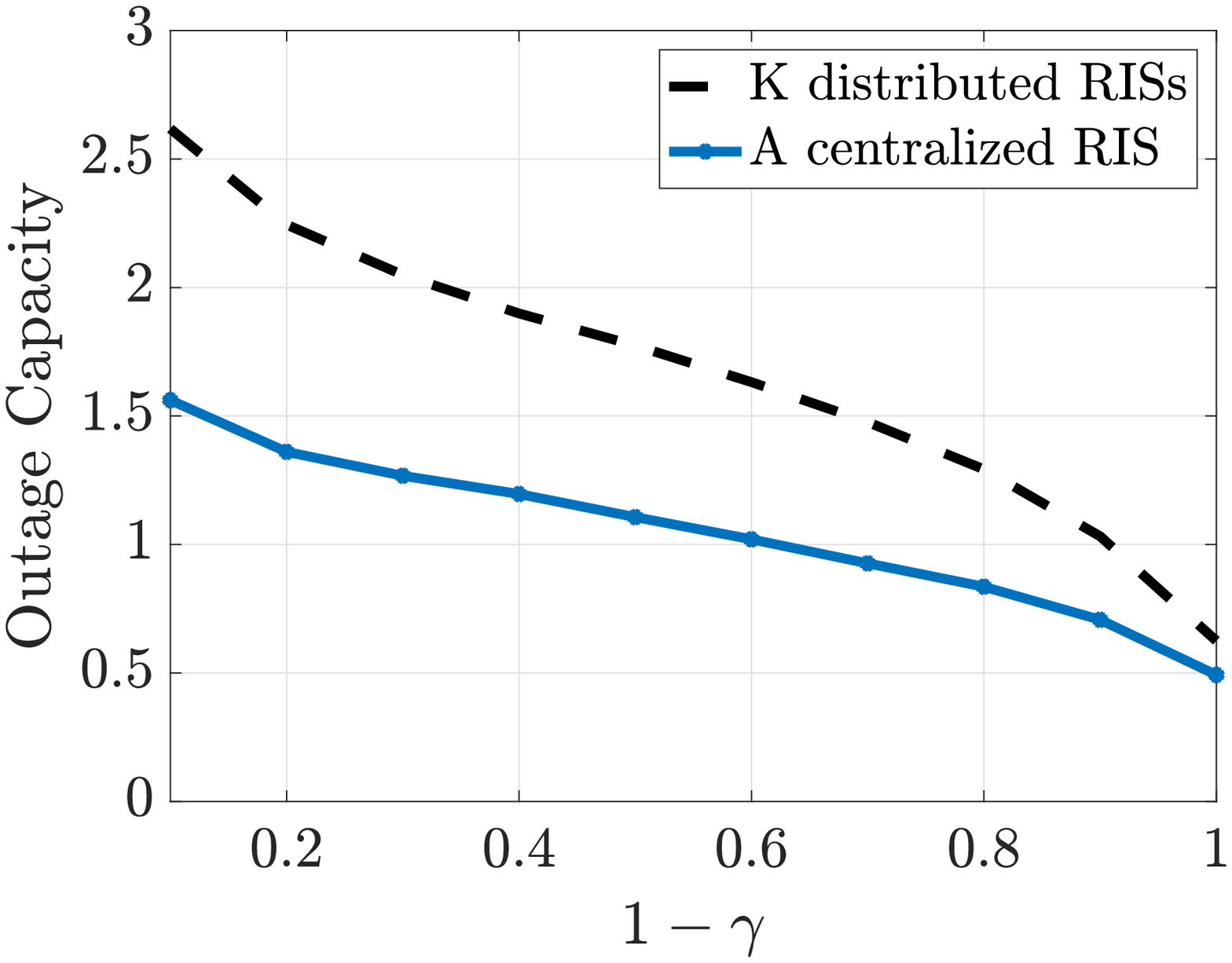}
	\end{minipage}
\vspace{-5mm}	
	
	\tikzset{every picture/.style={line width=0.75pt}} 

\begin{tikzpicture}[x=0.75pt,y=0.75pt,yscale=-1,xscale=1]

\draw (95.5,225) node [anchor=north west][inner sep=0.75pt]   [align=left] {\hspace{27.5mm}(a)\hspace{49mm}(b)\hspace{50mm}(c)};

\end{tikzpicture}
\vspace{-5mm}
\caption{\it (a) A comparison between the distributed RISs and three different scenarios based on the centralized RIS when $K=4$, $P_i=20$dB, and $M\in \{16,32,64,128\}$; (b) Sum-rate versus $P_i$ using the distributed RISs and different strategies based on a centralized RIS when $K=4$ and $M=32$; (c) Outage capacity analysis of using a centralized RIS and $K$ distributed RISs when $K=4$ 
  and the channels are noiseless. }\label{Fig:dist_vs_center_sum_rate}
\end{figure}

\textbf{Sum-rate analysis:} We investigate the performance of the distributed RISs with three distinct scenarios under a centralized RIS with noiseless channels. We regard $M$ as the total budget of the smart surfaces, meaning that $\text{RIS}_i, i = 1,2,\ldots, K$ contains $\frac{M}{K}$ elements and the centralized RIS utilizes $M$ elements. Suppose $K = 4$, $N=4$, $C_0=-30$ dB, the transmitters are available at $(50,0), (0,50), (100,50)$, and $(50,100)$, the distributed RISs are located at $(47,4), (3,54), (97,46),$ and $(53,96)$, and the receivers are spread at random in a room of size $100\times 100$. We assume three scenarios for the location of the centralized RIS: (1) Euclidean distance between transmitters (i.e., $(50,50)$), (2) near one of the transmitters (e.g., close to ${\sf Tx}_1$ at $(47,4)$), and (3) at random. We consider \eqref{eq:opt_min_one_RIS} and \eqref{eq:opt_max_min_score} as the optimization problems with a centralized RIS and the distributed RISs, respectively. In Fig.~\ref{Fig:dist_vs_center_sum_rate}(a), we show sum-rate against $\log(M)$ when $P_i = 20$dBm for $i=1,2,\ldots,K$. As Fig.~\ref{Fig:dist_vs_center_sum_rate}(a) shows, the distributed case outperforms all scenarios with a centralized RIS and needs fewer elements to attain a given sum-rate. This occurs since determining the optimal placement of the single RIS is challenging, whereas each distributed RIS is located near its associated transmitter. 
In addition, in Fig.~\ref{Fig:dist_vs_center_sum_rate}(b), we compare the aforementioned cases when $M=32$, $P_i \in [0,30]$ dBm at all transmitters, and all other assumptions are the same as in Fig.~\ref{Fig:dist_vs_center_sum_rate}(a). It depicts that the distributed scenario outperforms the other cases, and the curves become saturated at high transmit powers.

\textbf{Outage capacity analysis:} In this work, we mainly focus on the average rates. However, from a practical standpoint, it is important to understand the outage probability resulting in failure to achieve target rates. Thus, we study the RIS-assisted networks' outage capacity with distributed and centralized RISs under noiseless channels. We define $(1-\gamma)$ outage capacity for  the $i^{\mathrm{th}}$ user as 
\begin{align}
    C^{1-\gamma}_{i}=\{R_i: \Pr(R_i\geq R_0)\geq 1-\gamma\}, \quad i\in\{1,2,\ldots,K\},
\end{align}
where $R_i$ shows the rate for the $i^{\mathrm{th}}$ user in which there exists a configuration at RIS(s) with
channel capacity more than or equal to $R_0$ with a probability greater than or equal to $1-\gamma$. Fig.~\ref{Fig:dist_vs_center_sum_rate}(c) shows the outage capacity of the centralized and distributed scenarios. In the centralized case, we assume $50$, $5$, and $47.17$ represent the distance between ${\sf Tx}_i$ and ${\sf Rx}_j, i,j=\{1,2,3,4\}$, the distance between ${\sf Tx}_i$ and the RIS, and the distance between the RIS and ${\sf Rx}_j$, respectively. Then, in the distributed case, we consider the RISs are located
at $(25,25), (25,75), (75,75)$, and $(75,25)$. Here, we assume other assumptions are similar to Fig.~\ref{Fig:dist_vs_center_sum_rate}(b). Fig.~\ref{Fig:dist_vs_center_sum_rate}(c) shows that the distributed case provides a higher outage capacity than the centralized case, which confirms the conclusion in Fig~\ref{Fig:dist_vs_center_sum_rate}(a) and (b) that the distributed scenario outperforms the centralized scenario. 

\subsection{Efficiency of our optimization method}
\label{subsection:efficiency_our_opt_method}

In~\cite{wu2019beamforming,abdullah2022low}, the authors propose two SR-based optimization methods, which optimize the RIS elements in an iterative fashion. In this part, we compare our filled function method with these baselines in terms of sum-rate, minimum rate (min rate), and complexity. Furthermore, since the optimization problems in~\eqref{eq:opt_min_one_RIS} and \eqref{eq:opt_min_one_RIS_fairness} are NP-hard problems, it is fair to utilize two other benchmarks based on heuristic optimization methods as the SES method and GA. Here, we focus on the centralized scenario and assume that $\textbf{h}^{[ij]}$ is distributed via Rician fading with Rician factor $\kappa=2$.

\begin{figure}[ht]
\vspace{-10mm}
	\begin{minipage}[c][1\width]{
	   0.5\textwidth}
	   \centering
	   \includegraphics[width=1\textwidth]{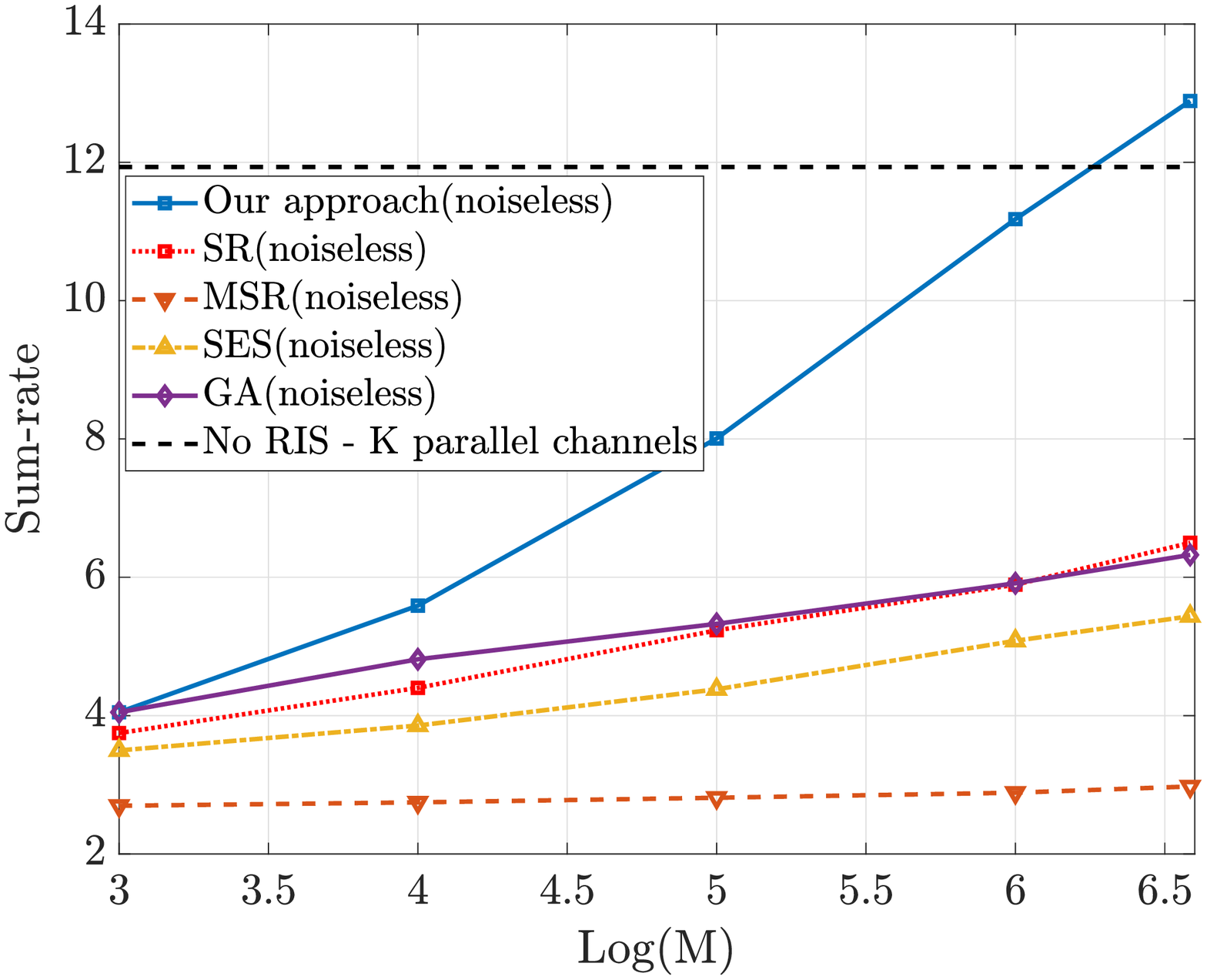}
	\end{minipage}
 \hfill 	
	\begin{minipage}[c][1\width]{
	   0.5\textwidth}
	   \centering
	   \includegraphics[width=1\textwidth]{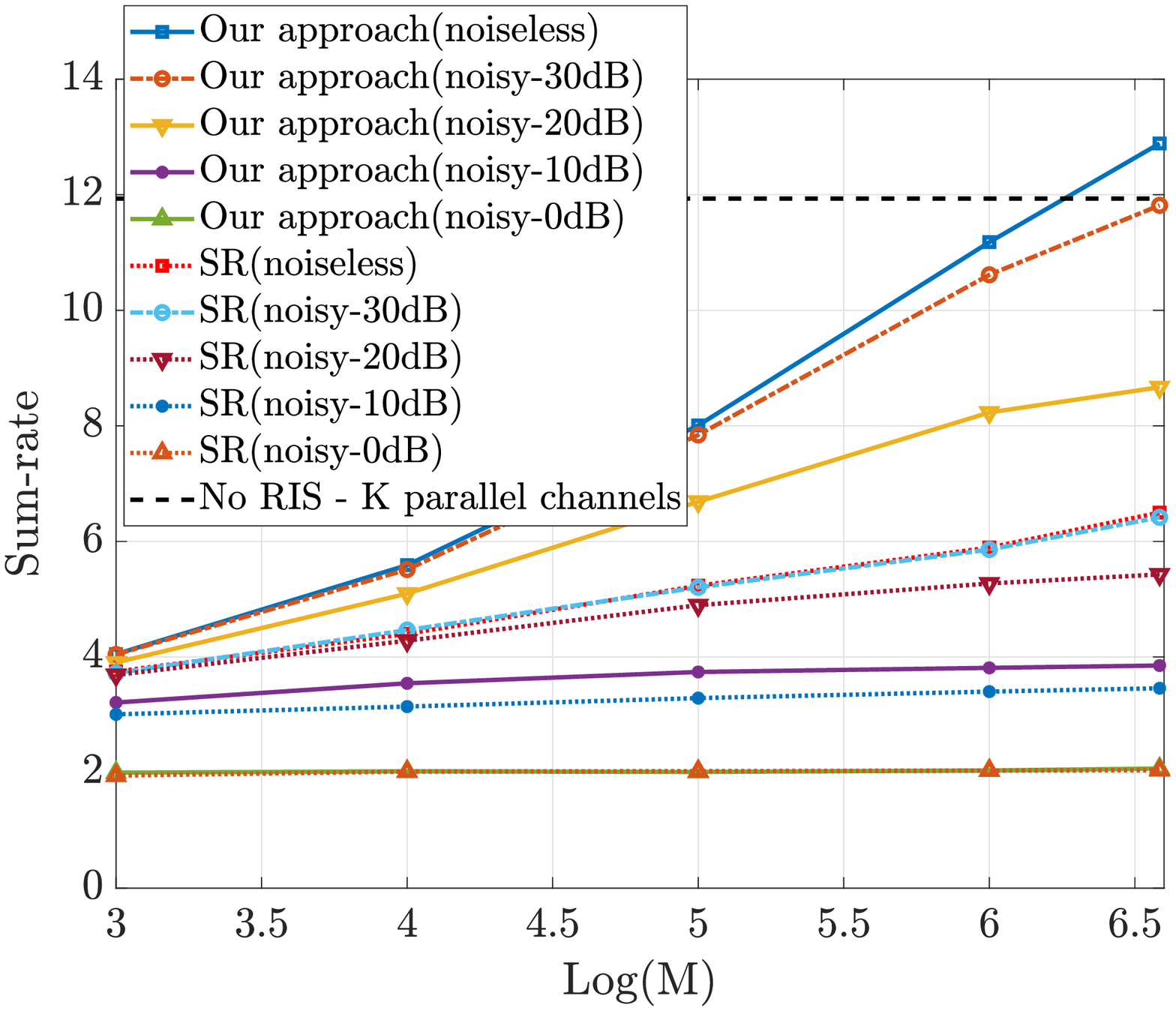}
	\end{minipage}
\vspace{-12mm}	
	\tikzset{every picture/.style={line width=0.75pt}} 

\begin{tikzpicture}[x=0.75pt,y=0.75pt,yscale=-1,xscale=1]

\draw (95.5,225) node [anchor=north west][inner sep=0.75pt]   [align=left] {\hspace{40.5mm}(a)\hspace{79mm}(b)};

\end{tikzpicture}
\vspace{-5mm}
\caption{\it (a) Sum-rate against $\log(M)$ of our approach, the SR method, the M-SR method, the SES method, GA, and No RIS - K parallel channels under noiseless channels when $K=4$ and $N=4$; (b) Sum-rate versus different values of $M$ using our approach and the SR method with different versions of the noisy channels. }\label{Fig:sum_rate_noiseless_noisy}
\end{figure}

\textbf{Sum-rate:} 
In Fig.~\ref{Fig:sum_rate_noiseless_noisy}(a), we assume $K = 4$, $N=4$, and $P_i=20$dBm for $ i=1,2,\ldots,K$. We also consider the channels are noiseless, $M\in \{8,16,32,64,96\}$, $C_0=-30$ dB, and $(0,0)$, $(50,0)$, and $(3,4)$ show the locations of the BSs, receivers, and RIS, respectively. Fig.~\ref{Fig:sum_rate_noiseless_noisy}(a) shows that our approach outperforms the SR-based methods as well as two heuristic optimization methods. Further, we consider a baseline with no RIS, referred to as ``No RIS-$K$ parallel channels,'' where each transmitter delivers its message to its desired receiver with no interference. Fig.~\ref{Fig:sum_rate_noiseless_noisy}(a) shows that our method offers a higher sum-rate than this baseline when $M=96$. The gap between our approach and the benchmarks grows as $M$ increases since our approach finds an approximate global solution while the others get stuck in the local minimizers. Then, in Fig.~\ref{Fig:sum_rate_noiseless_noisy}(b), we present the sum-rate of our approach and the SR method using both noisy and noiseless channels. It depicts that our optimization approach offers acceptable sum-rate values using noisy-$30$ and noisy-$20$ channels; however, it fails to perform well when we utilize noisy-$10$ and noisy-$0$ channels. It occurs because, as the noise power in the noisy CSI grows, the impact of using more RIS elements diminishes since the actual CSI is substantially different from the noisy CSI used in the optimization process.

\begin{figure}[ht]
\vspace{-10mm}
	\begin{minipage}[c][1\width]{
	   0.5\textwidth}
	   \centering
	   \includegraphics[width=1\textwidth]{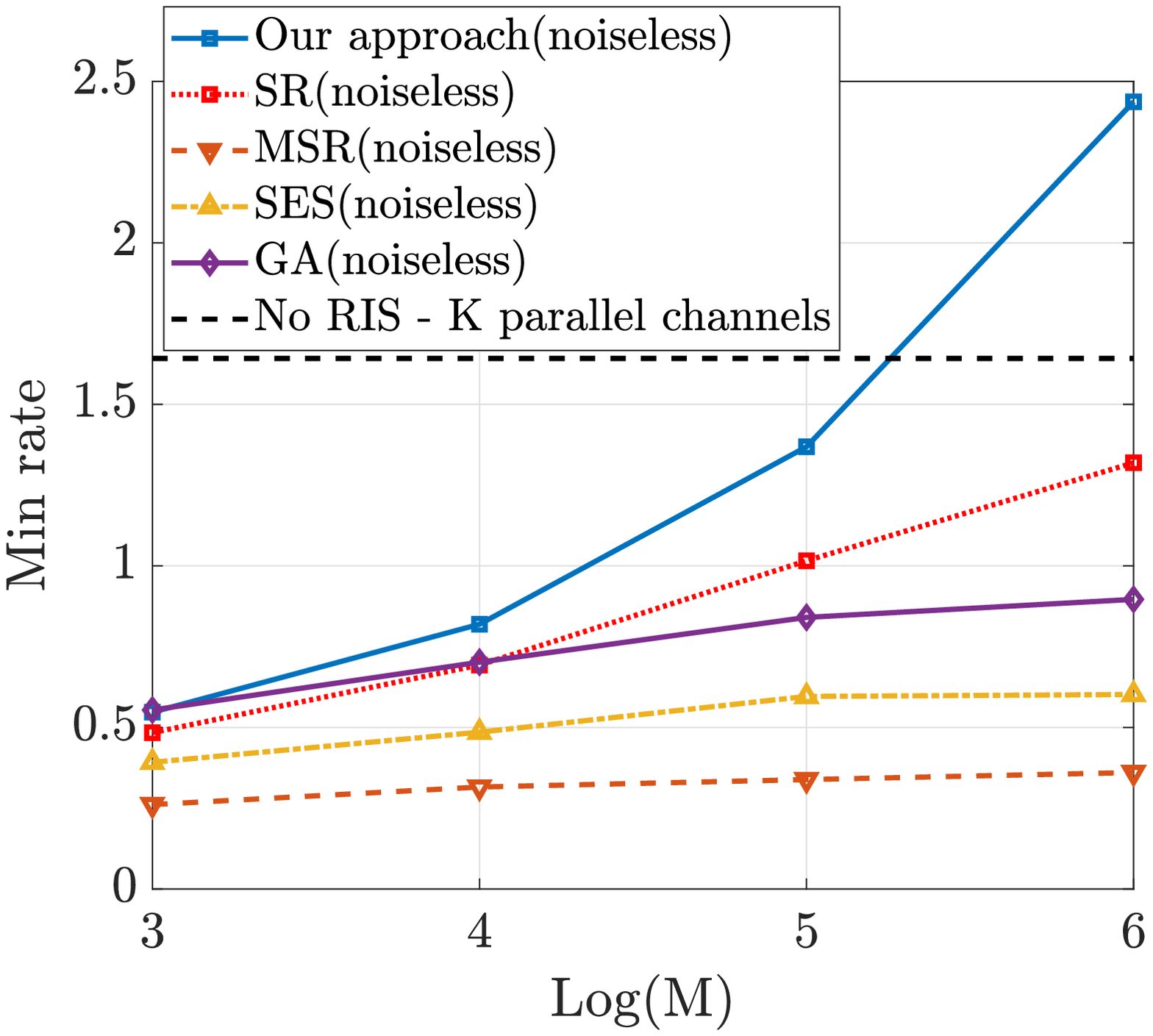}
	\end{minipage}
 \hfill 	
	\begin{minipage}[c][1\width]{
	   0.5\textwidth}
	   \centering
	   \includegraphics[width=1\textwidth]{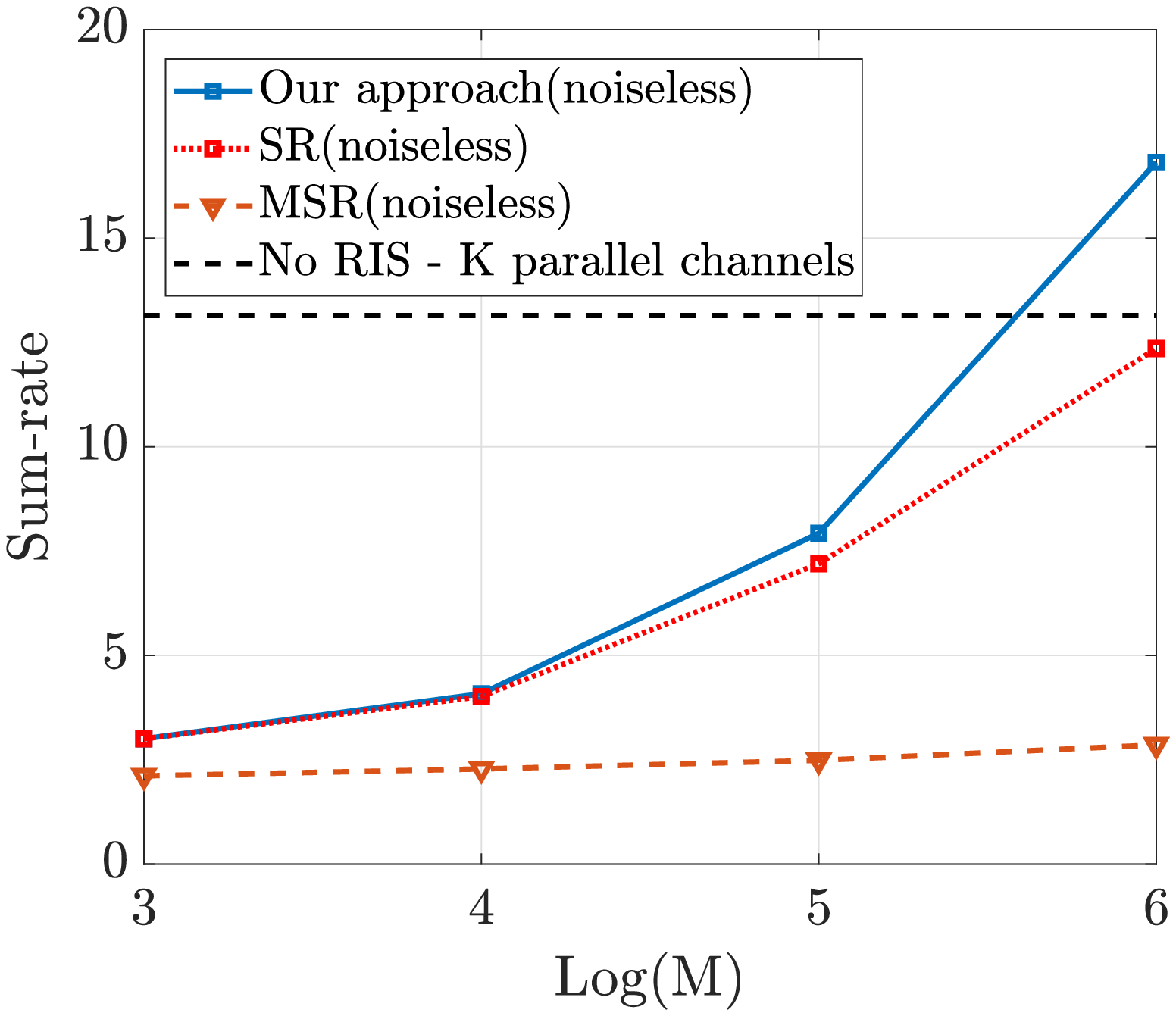}
	\end{minipage}
\vspace{-10mm}	
	\tikzset{every picture/.style={line width=0.75pt}} 

\begin{tikzpicture}[x=0.75pt,y=0.75pt,yscale=-1,xscale=1]

\draw (95.5,225) node [anchor=north west][inner sep=0.75pt]   [align=left] {\hspace{40.5mm}(a)\hspace{79mm}(b)};

\end{tikzpicture}
\vspace{-5mm}
\caption{\it (a) Min rate versus $\log(M)$ of our approach, the SR method, the M-SR method, the SES method, GA, and No RIS - K parallel channels under noiseless channels; (b) A comparison between our approach, the SR method, and the M-SR method using Nakagami fading when $K=4$, and $M\in\{8,16,32,64\}$ }\label{Fig:min_rate_noiseless}
\end{figure}

\textbf{Min-rate:} 
To ensure user fairness, we apply the max-min optimization problem in \eqref{eq:opt_min_one_RIS_fairness} and compute the min rate. The assumptions are analogous to those in Fig.~\ref{Fig:sum_rate_noiseless_noisy}(a). As shown in Fig.~\ref{Fig:min_rate_noiseless}(a), our approach provides appreciable gains over the SR-based methods as well as GA and the SES approach due to its more efficient optimization technique. Additionally, when $M=64$, Fig.~\ref{Fig:min_rate_noiseless}(a) indicates that our approach achieves a higher min rate than the No RIS-$K$ parallel channels.

\textbf{Complexity:} In this part, we compute the complexity of our approach, the brute force search, and the baselines described in Fig.~\ref{Fig:sum_rate_noiseless_noisy}(a). We use the number of evaluations required to obtain the solution as a proxy for complexity, which can be derived as follows. 
\begin{align}
\label{eq:complexity}
    &\left(N-1 \right)M i_{0}^{\text{loc}}+\sum_{m=0}^{\left(N-1 \right)M}\left(\frac{\left(N-1 \right)M i_{\ell}^{\text{loc}}}{\tau}+\left(N-1 \right)M i_{\ell}^{\text{loc}}\right)i_{m}^{\text{filled}} \nonumber \\
    &\overset{(a)}{\leq} \left(N-1 \right)M i_{\max}^{\text{loc}}+\frac{\tau+1}{\tau}\left(N-1 \right)Mi_{\max}^{\text{loc}}\underbrace{i_{\max}^{\text{filled}}}_{\overset{(b)}{\leq}N\left((N-1)M+1\right)\left[\log_{10}(r/\epsilon)+1\right]} \nonumber \\
    &\leq \left(N-1 \right)M i_{\max}^{\text{loc}}+\frac{\tau+1}{\tau}N\left(N-1 \right)M\left((N-1)M+1\right)\left[\log_{10}(r/\epsilon)+1\right]\underbrace{i_{\max}^{\text{loc}}}_{\overset{(c)}{\leq}M}  \nonumber\\
    &\equiv \mathcal{O}\left(N\left(N-1 \right)^2 \log_{10}(r/\epsilon)M^3\right),
\end{align}
where $(a)$ happens since $i_{\ell}^{\text{loc}}\leq i_{\max}^{\text{loc}}$ and $\sum_{m=0}^{\left(N-1 \right)M} i_{m}^{\text{filled}} \leq i_{\max}^{\text{filled}}$, and $(b)$ and $(c)$ follow the definition of Algorithms~\ref{algo:Local} (line $7$) and \ref{algo:global} (line $25$) where $i_{\max}^{\text{loc}}$ and  $i_{\max}^{\text{filled}}$ cannot be greater than $M$ and $N\left((N-1)M+1\right)$ $\left[\log_{10}(r/\epsilon)+1\right]$, respectively. 
In \eqref{eq:complexity}, we use the Landau notation (``big
O'') in its standard form, and $\log_{10}$ represents the logarithmic function in base $10$. 

We use the same assumptions as in Fig.~\ref{Fig:sum_rate_noiseless_noisy}(a) and calculate the complexity of our method, the brute force search, the SR-based methods, GA, and the SES method when $M \in \{32,64\}$ as Table~\ref{table:compare_complexity}. This table shows that our method is faster than the brute-force search and the SES method but slower than the others, revealing the rate-complexity trade-off of the methods. The results of Table~\ref{table:compare_complexity} support the analytical complexity statement in \eqref{eq:complexity}.

In this work, we aim to attain the optimal RIS configuration with negligible overhead. For instance, at carrier frequency $f_c=1.8$GHz and at the speed of $v=60\mathrm{km/h}$, the channel coherence time is equal to $c/(f_{c}v)=10\mathrm{msec}$, where $c$ denotes the speed of light. According to Table~\ref{table:compare_complexity}, our method requires $1.45\times 10^6$ evaluations to find the optimal solution with $M=64$, and if we assume a dual-core processor with a clock frequency of $2$ GHz, our approach requires approximately $0.36\mathrm{msec}$ to get the results, which is negligible.
\begin{table}[hb]
\caption{The complexity of our approach, brute force search, and the benchmarks mentioned in section~\ref{subsection:efficiency_our_opt_method}.
}
\centering
\begin{tabular}{|c|c|c|c|c|c|}
\hline
Method & $M=32$ & $M=64$ & Method & $M=32$ & $M=64$\\

\hline \hline
Brute force & $1.84\times 10^{19}$ &  $3.40\times10^{38}$ & SR method&  $2.34\times10^4$ &  $4.71\times10^{4}$ \\
M-SR method & $1.28\times10^4$ & $2.56\times10^{4}$ & Genetic algorithm &  $2.01\times10^4$ &  $2.01\times10^{4}$  \\
Simplified exhaustive search& $3.17\times10^5$ & $2.67\times10^{6}$ & \textbf{Sigmoid filled function} &  $3.63\times 10^5$ & $1.45\times 10^6$\\
\hline
\end{tabular}
\label{table:compare_complexity}
\end{table}

\subsection{Nakagami channel model}
In this part, we use a more realistic channel model based on Nakagami distribution in~\cite{selimis2021performance}. Specifically, we define $h_m^{[j]}=|h_m^{[j]}|e^{j\bar{\theta}_{h,m}^{[j]}}$, $g_m^{[i]}=|g_m^{[i]}|e^{j\bar{\theta}_{g,m}^{[i]}}$, and $h_d^{[ji]}=|h_d^{[ji]}|e^{j\bar{\theta}_d^{[ji]}}$ as the communication channels, 
where $\bar{\theta}_{h,m}^{[j]}$, $\bar{\theta}_{g,m}^{[i]}$, and $\bar{\theta}_d^{[ji]}$ describe the angles of $h_m^{[j]}$, $g_m^{[i]}$, and $h_d^{[ji]}$, respectively. Here, $|h_m^{[j]}|$, $|g_m^{[i]}|$, and $|h_d^{[ji]}|$ follow a Nakagami distribution with parameters $(m_{h_m^{[j]}}, \Omega_{h_m^{[j]}})$, $(m_{h_m^{[i]}}, \Omega_{g_m^{[i]}})$, and $(m_{h_d^{[ji]}}, \Omega_{h_d^{[ji]}})$, respectively, where $\Omega$ parameters represent the large scale fading of the channels. Moreover, the angles of the channels follow independent uniform distributions. We use the cosine law to take the angle of the incident and the reflected signals at RIS into consideration. More precisely, we assume 
\begin{align}
    d_{{\sf Tx}_i-{\sf Rx}_i}=\sqrt{d^2_{{\sf Tx}_i-\text{RIS}^{[m]}}+d^2_{\text{RIS}^{[m]}-{\sf Rx}_i}-2d_{{\sf Tx}_i-\text{RIS}^{[m]}}d_{\text{RIS}^{[m]}-{\sf Rx}_i}\cos\left(\psi_m\right)},
\end{align}
where $d_{{\sf Tx}_i-\text{RIS}^{[m]}}$, $d_{\text{RIS}^{[m]}-{\sf Rx}_i}$, and $d_{{\sf Tx}_i-{\sf Rx}_i}$ indicate the distance between ${\sf Tx}_i$ and the $m^{\mathrm{th}}$ element of the RIS, between the $m^{\mathrm{th}}$ element of the RIS and ${\sf Rx}_i$, and between ${\sf Tx}_i$ and ${\sf Rx}_i$, respectively. We use $\psi_m$ to denote the angle between the ${\sf Tx}_i-\text{RIS}^{[m]}$ and $\text{RIS}^{[m]}-{\sf Rx}_i$ links. Fig.~\ref{Fig:min_rate_noiseless}(b) depicts the comparison results between our approach and two SR-based methods using a centralized RIS when the channels are noiseless, $K = 4$, $N=4$, and $P_i=20$dBm for $ i=1,2,\ldots,K$. We assume $M\in \{8,16,32,64\}$, $m_{h_d^{[ji]}}=3$, $m_{h_m^{[j]}}=1.5$, $m_{g_m^{[i]}}=2.5$, $\psi_m=86^{\circ}$, $C_0=-31.5$dB, and $(0,0)$, $(50,0)$, and $(3,4)$ are the locations of the BSs, receivers, and RIS, respectively. According to Fig.~\ref{Fig:min_rate_noiseless}(b), our filled function-based method provides a higher sum-rate than the other baselines due to finding the approximation of the global solutions instead of the local optimal solutions.

\subsection{Minimum required degrees-of-control for the RIS }
\label{subsection:simulation_lower_bound}
In this section, we analyze the minimum number
of required RIS elements to obtain a desired performance
metric. In particular, we focus on the following cases.

\begin{figure}[ht]
\vspace{-12mm}
	\begin{minipage}[c][1\width]{
	   0.5\textwidth}
	   \centering
	   \includegraphics[width=1\textwidth]{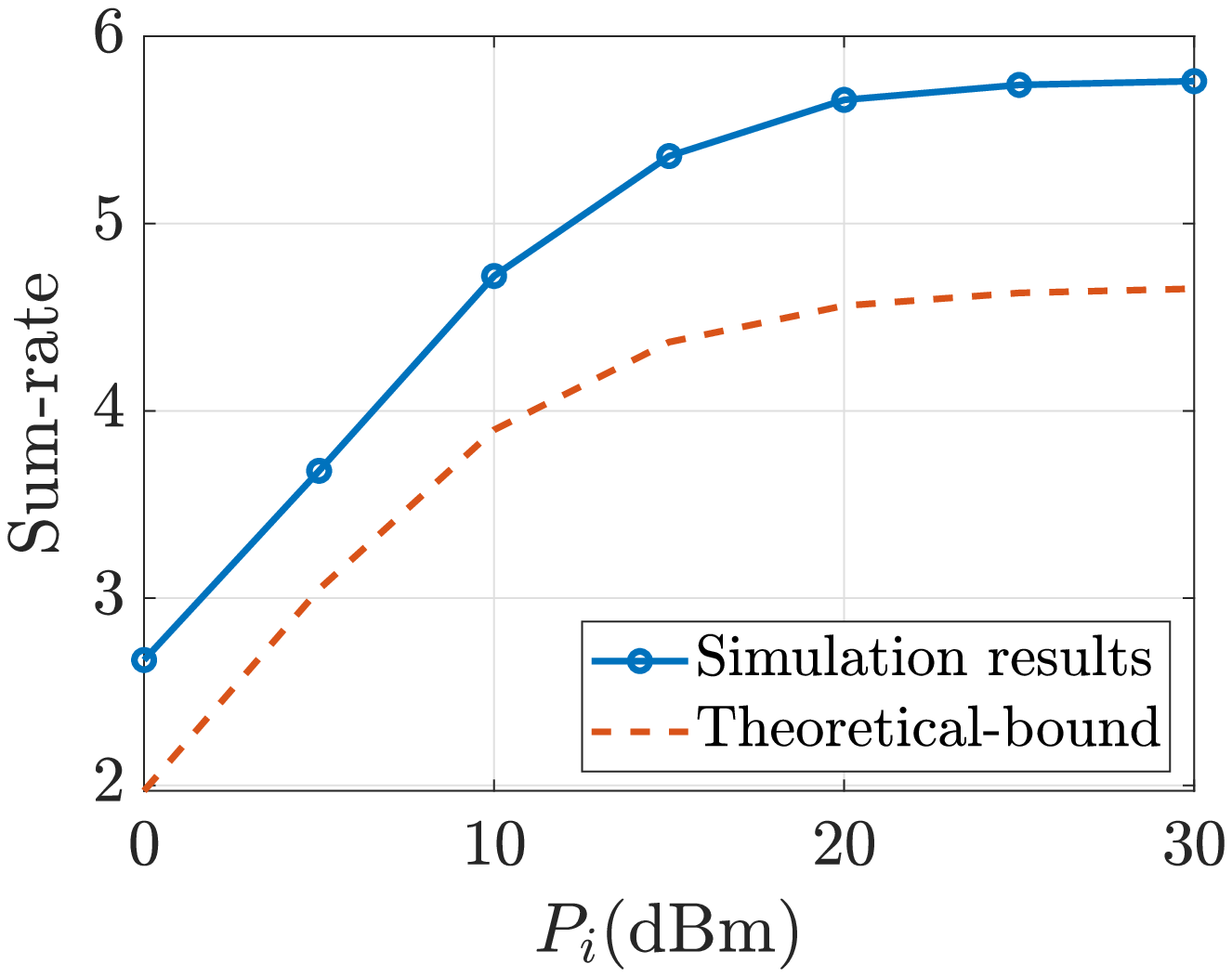}
	\end{minipage}
 \hfill 	
	\begin{minipage}[c][1\width]{
	   0.5\textwidth}
	   \centering
	   \includegraphics[width=1\textwidth]{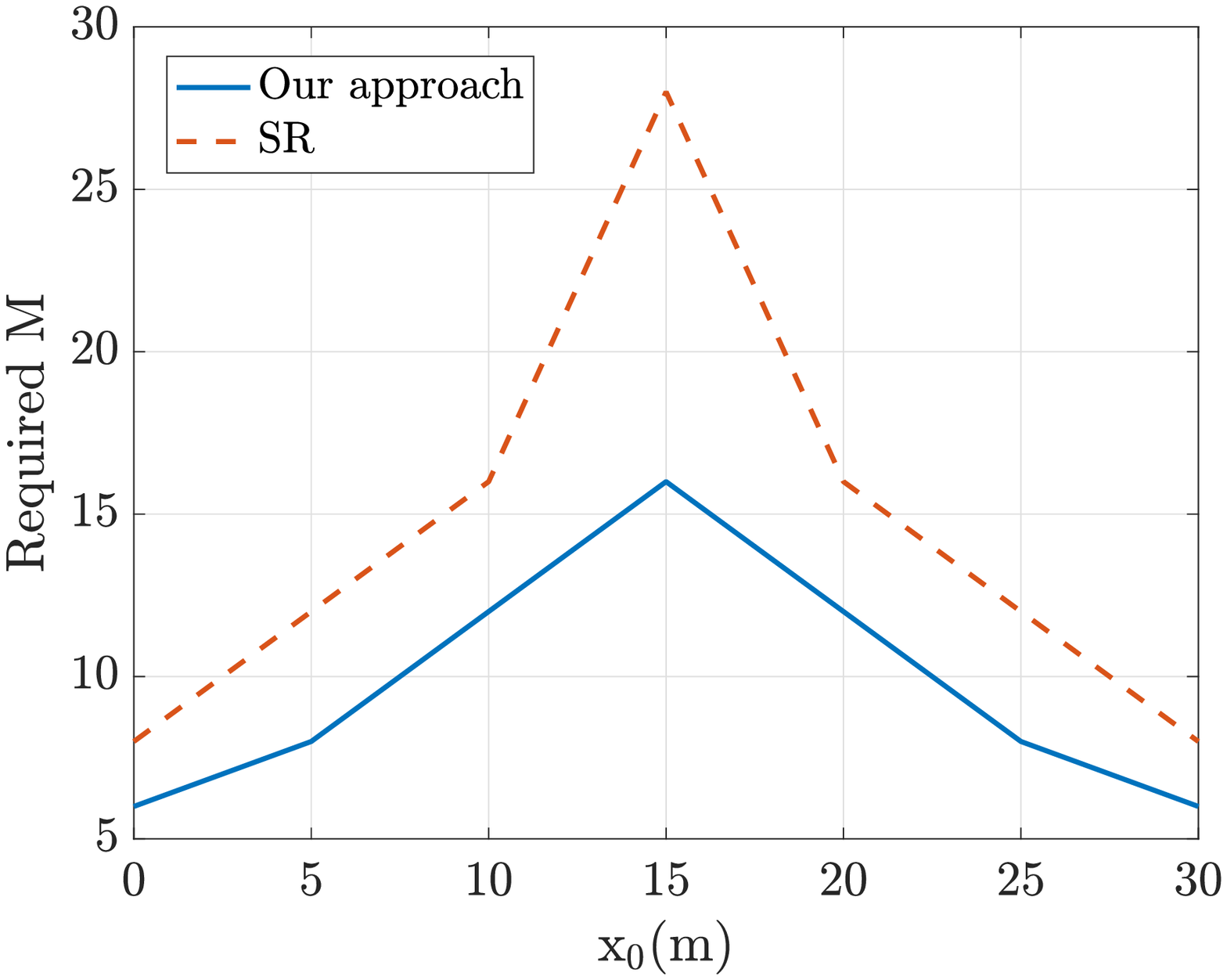}
	\end{minipage}
\vspace{-10mm}	
	\tikzset{every picture/.style={line width=0.75pt}} 

\begin{tikzpicture}[x=0.75pt,y=0.75pt,yscale=-1,xscale=1]

\draw (95.5,225) node [anchor=north west][inner sep=0.75pt]   [align=left] {\hspace{40.5mm}(a)\hspace{79mm}(b)};

\end{tikzpicture}
\vspace{-5mm}
\caption{\it (a) A comparison between our approach and the theoretical-bound with a centralized RIS when $M_{\min}=24$, $K=3$; (b) The number of required RIS elements to have the sum-rate of $4$ with one RIS when $P_i=30$dBm, $K=4$, and the channels are noisy-$30$.}\label{Fig:lower_bound}
\end{figure}

\textbf{Theoretical lower-bound:} We assume a centralized RIS with noiseless channels and consider a symmetric setting where $K=3$, $N=8$, the path loss exponent between each pair of transmitter-receiver is equal to $3.9$, the distance between each pair of transmitter-receiver, the distance between each transmitter and the $\text{RIS}$ and the distance between the $\text{RIS}$ and each receiver are equal to $25, \sqrt{2}$, and $24.02$, respectively. To obtain an accurate curve, we solve \eqref{eq:final_lower_bound_equation} numerically without utilizing the practical interval for $M$. In Fig.~\ref{Fig:lower_bound}(a), we plot sum-rate versus $P_i$ with the theoretical-bound and our approach when $M_{\min}=24$. The results show that our method provides higher results than the theoretical-bound since we obtain the theoretical-bound pessimistically (i.e., the RIS configurations are independent of the channels).

\textbf{RIS at different locations:} To study the optimal placement of the RIS, we calculate the minimum number of the RIS elements that our method and the SR method require to provide a certain 
sum-rate when the locations of ${\sf Tx}_i$, the RIS, and ${\sf Rx}_i$ are equal to $(0,0)$, $(x_0,1), x_0\in[0,30]$, and $(30,0)$, respectively. Moreover, we assume a symmetric setting with a centralized RIS where $P_i=30$dBm, the channels are noisy-$30$, $K = 4$, $N=4$, and $C_0=-30$ dB. Then, we compute the number of required RIS elements to achieve a sum-rate of $4$ as described in Fig.~\ref{Fig:lower_bound}(b). Not surprisingly, due to the significant product path-loss, the number of required elements is highest when the RIS is placed in the middle between the transmitters and receivers, and the RIS needs fewer elements as it moves closer to either the transmitters or the receivers. In addition, Fig.~\ref{Fig:lower_bound}(b) describes that our method requires fewer elements than the SR method due to its better optimization performance.

\section{Conclusion}
\label{Section:conclusion}
In this paper, we investigated a SISO IoT RIS-assisted network under (1) centralized RIS and (2) distributed RISs, with one RIS allocated to each transmitter. The simulation results demonstrated that the distributed scenario offers a higher sum-rate than the centralized case and requires fewer RIS elements to provide the same performance. We proposed an optimization approach based on a sigmoid filled function to optimize the RIS elements in the discrete domain. We showed that our optimization approach provides a higher rate and requires fewer RIS elements to meet a certain sum-rate, compared to the SR-based methods~\cite{wu2019beamforming,abdullah2022low}, GA, and SES methods. Finally, we evaluated the minimum required degrees-of-control for the RIS to obtain a desired
performance metric. A future direction for this work would assume a wider range for the number of distributed RISs. 
The main challenges are the placement of the RIS and pairing them with the transmitters.

\begin{appendices}
\section{Proof of Lemma~\ref{lemma:approx_prob}}

\label{appndx:proof_approx_prob}

\begin{proof}
For a given $\mathbf{\Theta}$, we have
\begin{align}
\label{eq:intersect_real_imag}
    &\Pr\Big(\Big|\textbf{g}^{[i]}\mathbf{\Theta}\textbf{h}^{[j]}\Big|<\frac{\delta_{ji}}{2}\Big|\mathbf{\Theta}\Big) >\Pr\Big(\Big|Re\{\textbf{g}^{[i]}\mathbf{\Theta}\textbf{h}^{[j]}\}\Big|<\frac{\delta_{ji}}{2\sqrt{2}}\Big|\mathbf{\Theta} \bigcap \Big|Im\{\textbf{g}^{[i]}\mathbf{\Theta}\textbf{h}^{[j]}\}\Big|<\frac{\delta_{ji}}{2\sqrt{2}}\Big|\mathbf{\Theta}\Big).
\end{align}

To simplify \eqref{eq:intersect_real_imag}, we need to show that $\Big|Re\{\textbf{g}^{[i]}\mathbf{\Theta}\textbf{h}^{[j]}\}\Big|$ and $\Big|Im\{\textbf{g}^{[i]}\mathbf{\Theta}\textbf{h}^{[j]}\}\Big|$ are independent. Notice that the independence of the real and imaginary parts of $\textbf{g}^{[i]}\mathbf{\Theta}\textbf{h}^{[j]}$ leads to the independence of $\Big|Re\{\textbf{g}^{[i]}\mathbf{\Theta}\textbf{h}^{[j]}\}\Big|$ and $\Big|Im\{\textbf{g}^{[i]}\mathbf{\Theta}\textbf{h}^{[j]}\}\Big|$. Based on Lemma~\ref{lemma:avg_variance}, $Re\{\textbf{g}^{[i]}\mathbf{\Theta}\textbf{h}^{[j]}\}$ and $Im\{\textbf{g}^{[i]}\mathbf{\Theta}\textbf{h}^{[j]}\}$ are distributed via zero-mean Gaussian distribution; therefore, $Re\{\textbf{g}^{[i]}\mathbf{\Theta}\textbf{h}^{[j]}\}$ and $Im\{\textbf{g}^{[i]}\mathbf{\Theta}\textbf{h}^{[j]}\}$ are independent if they are uncorrelated (i.e., their covariance is equal to zero), which is straightforward. 

Then, for $i,j\in\{1,2,\ldots,K\}, j\neq i$, we simplify \eqref{eq:intersect_real_imag} as 
\begin{align}
\label{eq:prob_real_imag}
    &\Pr\Big(\Big|\textbf{g}^{[i]}\mathbf{\Theta}\textbf{h}^{[j]}\Big|<\frac{\delta_{ji}}{2}\Big|\mathbf{\Theta}\Big) >\Pr\Big(\Big|Re\{\textbf{g}^{[i]}\mathbf{\Theta}\textbf{h}^{[j]}\}\Big|<\frac{\delta_{ji}}{2\sqrt{2}}\Big|\mathbf{\Theta} \Big) \Pr\Big( \Big|Im\{\textbf{g}^{[i]}\mathbf{\Theta}\textbf{h}^{[j]}\}\Big|<\frac{\delta_{ji}}{2\sqrt{2}}\Big|\mathbf{\Theta}\Big).
\end{align}

Since $Re\{\textbf{g}^{[i]}\mathbf{\Theta}\textbf{h}^{[j]}\}$ is a zero-mean Gaussian random variable, we have
\begin{align}
\label{eq:prob_real}
    &\Pr\Big(\Big|Re\{\textbf{g}^{[i]}\mathbf{\Theta}\textbf{h}^{[j]}\}\Big|<\frac{\delta_{ji}}{2\sqrt{2}}\Big|\mathbf{\Theta}\Big)=\int_{-\frac{\delta_{ji}}{2\sqrt{2}}}^{\frac{\delta_{ji}}{2\sqrt{2}}}\frac{1}{\sqrt{\pi \nu}}e^{\frac{-r^2}{\nu}}dr \overset{(a)}{\approx}\int_{-\frac{\delta_{ji}}{2\sqrt{2}}}^{\frac{\delta_{ji}}{2\sqrt{2}}}\frac{1}{\sqrt{\pi \nu}}\Big[1-\frac{r^2}{\nu}+O (\frac{r^4}{\nu^2})\Big]dr \nonumber \\
    & \quad \quad \quad \quad =\frac{1}{\sqrt{\pi \nu}} \Big[\frac{\delta_{ji}}{\sqrt{2}}-\frac{\delta_{ji}^3}{24\sqrt{2}\nu}+O(\frac{\delta_{ji}^5}{\nu^2}) \Big]\overset{(b)}{\approx}\frac{\delta_{ji}}{\sqrt{2\pi \nu}},~~ i,j\in\{1,2,\ldots,K\}, j\neq i,
\end{align}
where $(a)$ applies based on approximation from the Taylor expansion of $e^{\frac{-r^2}{\nu}}$, and $(b)$ holds since our goal is to have a small $\delta_{ji}$; therefore, we ignore the high orders of $\frac{\delta_{ji}}{\nu}$.

Similarly, we obtain
\begin{align}
\label{eq:prob_imag}
    \Pr\Big(\Big|Im\{\textbf{g}^{[i]}\mathbf{\Theta}\textbf{h}^{[j]}\}\Big|<\frac{\delta_{ji}}{2\sqrt{2}}\Big|\mathbf{\Theta}\Big)\approx\frac{\delta_{ji}}{\sqrt{2\pi \nu}}, ~~ i,j\in\{1,2,\ldots,K\}, j\neq i.
\end{align}

Hence, using \eqref{eq:prob_real} and \eqref{eq:prob_imag} leads to
\begin{align}
    \Pr\Big(\Big|\textbf{g}^{[i]}\mathbf{\Theta}\textbf{h}^{[j]}\Big|<\frac{\delta_{ji}}{2}\Big|\mathbf{\Theta}\Big)&>(\frac{\delta_{ji}}{\sqrt{2\pi \nu}})^2,~~ i,j\in\{1,2,\ldots,K\}, j\neq i,
\end{align}
which completes the proof.
\end{proof}
\end{appendices}

{\footnotesize
\bibliographystyle{ieeetr}
\bibliography{refs.bib}
}

\end{document}